\documentclass{aa} 
\usepackage{graphicx,lscape,natbib,epsfig}
\bibpunct{(}{)}{;}{a}{}{,} 
\begin{document}

\def\spk{$S_{2.7 \rm \thinspace GHz}$ }
\def\bj{$b_{J}$ }
\def\MHz{{\rm\thinspace MHz}}
\def\GHz{{\rm\thinspace GHz}}
\def\Sr{{\rm\thinspace Sr}}
\def\Jy{{\rm\thinspace Jy}}
\def\iraf{{\small IRAF}}
\def\ra{{\rm{RA}}}
\def\dec{{\rm{Dec}}}
\def\d{\hbox{$^\circ$}}
\def\h{\hbox{$^{\rm h}$}}
\def\eg{{\it e.g.}}

\def\apj{{ApJ}}  
\def\apjl{{ApJL}} 
\def\apjs{{ApJS}}  
\def\aj{{AJ}}
\def\pasp{{Pub.~Astron. Soc. Pacif.}} 
\def\mnras{{MNRAS}}      
\def\aap{{A\&A}}     
\def\aasup{{A\&AS}}     
\def\apss{{Ap\&SS}} 
\def\ajp{{Aust. J. Phys.}} 
\def\sci{{Science}}
\def\nat{{Nature}}
\def\physrep{{Phys. Rep.}}

\title{The Parkes quarter-Jansky flat-spectrum sample}

\subtitle{3. Space density and evolution of QSOs}

  \author{J.V. Wall
          \inst{1}\fnmsep\thanks{now at Department of Physics and Astronomy, University of British Columbia,
            Vancouver, B.C. V6T 1Z1, Canada}
          \and
          C.A. Jackson\inst{2}\fnmsep\thanks{now at Australia Telescope
            National Facility, CSIRO, PO Box 76, Epping, NSW 1710, Australia}
          \and
          P.A. Shaver\inst{3}
          \and
          I.M. Hook\inst{1}
          \and
          K.I. Kellermann\inst{4}
          }

   \institute{Department of Astrophysics, University of Oxford,
       Denys Wilkinson Building, Keble Road, Oxford OX1 3RH, UK
         \and
          Research School of Astronomy \& Astrophysics,
       The Australian National University, Mount Stromlo Observatory,
       Canberra, ACT 2611, Australia
          \and
        European Southern Observatory, Karl-Schwarzschild-Strasse 2,
               85748 Garching bei M\"unchen, Germany
          \and
National Radio Astronomy Observatory, 520 Edgemont Road,
        Charlottesville, VA 22903-2475, USA\\
             }


 \offprints{J.V. Wall, \email{jvw@astro.ubc.ca}}

   \date{2005 Jan 10}

\abstract{We analyze the Parkes quarter-Jansky flat-spectrum sample of QSOs in
terms of space density, including the redshift distribution, the radio
luminosity function, and the evidence for a redshift cutoff. With regard to the
luminosity function, we note the strong evolution in space density from the
present day to epochs corresponding to redshifts $\sim1$.  We draw attention to
a selection effect due to spread in spectral shape that may have misled other
investigators to consider the apparent similarities in shape of luminosity
functions in different redshift shells as evidence for luminosity evolution. To
examine the evolution at redshifts beyond $3$, we develop a model-independent
method based on the $V_{\rm max}$ test using each object to predict expectation
densities beyond $z=3$. With this we show that a diminution in space density at
$z > 3$ is present at  a significance level $>4\sigma$. We identify a severe
bias in such determinations from using flux-density measurements at epochs
significantly later than that of the finding survey. The form of the diminution
is estimated, and is shown to be very similar to that found for QSOs selected
in X-ray and optical wavebands. The diminution is also compared with the
current estimates of star-formation evolution, with less conclusive results. In
summary we suggest that the reionization epoch is little influenced by powerful
flat-spectrum QSOs, and that dust obscuration does not play a major role in our
view of the QSO population selected at radio, optical or X-ray wavelengths.
}

\maketitle
%

\section{Introduction}

This is the last in a series of three papers describing the results of a
program to search for high-redshift radio-loud QSOs and to study the evolution
of the flat-spectrum QSO population.

Paper 1 \citep{jac02} set out the sample, discussing selection, identification
and reconfirmation programmes to determine the optical counterparts to the
radio sources.  Paper~2 \citep{hoo03} presented new spectroscopic observations
and redshift determinations.  This paper considers the radio-loud QSO space
distribution, the epoch-dependent luminosity function, the evidence for a
redshift cutoff provided by the sample, and the form of this cutoff.

Paper 1 described how the identification programme for 878 flat-spectrum radio
sources selected from the Parkes catalogues yielded a near-complete set of
optical counterparts. Indeed for the sub-sample at declinations above $-$40\d\
with flux densities above catalogue completeness limits, only one source
remains unidentified. Of the 379 QSOs in this sub-sample, 355 have measured
redshifts, obtained from earlier observations and the extensive spectroscopy
programme described in Paper 2. This relative completeness is ideal for studies
of space density, as it becomes possible to map the entire ``quasar epoch''
with a single homogeneous sample, having no optical magnitude limit and free of
obscuration effects. In fact a sub-sample of objects from an earlier analysis
was used by \citet{sha96} to study the evolution of QSO space density at high
redshifts. The study showed that the space density of high-luminosity radio
QSOs decreased significantly at redshifts beyond 3. Preliminary data were also
used by \cite{jac99} in considering a dual-population scheme of space densities
for unified models of QSOs and radio galaxies.

General features of the luminosity function and its redshift dependence have
long been established for QSOs selected at optical and radio wavelengths (e.g.
\citealt{lon66,sch68,fan73}).  Powerful evolution is required, similar in
magnitude for selection at either waveband; the space density of the more
luminous QSOs at redshifts of 1 to 2 is at least $10^2$ that at the present
epoch. It has been hotly debated as to whether the form of this change is
luminosity evolution (e.g. \citealt{boy88}) or luminosity-dependent density
evolution (e.g. \citealt{dun90}). It does not matter: physical models are not
available that {\it require} either form, although it is clear that luminosity
evolution results in lifetimes of non-physical length \citep[ and references
therein]{hae93b}. The space density of radio-selected QSOs, constituting some
10 per cent of all QSOs, generally appears to parallel that of
optically-selected QSOs ({\it e.g.} \citealt{sch91,ste00}).

There are many reports of a redshift cutoff in the literature: paper after
paper speaks of `the quasar epoch', `a strongly-evolving population peaking at
a redshift of about 2', or `the quasar redshift cutoff' without specific
reference.  For optically-selected QSOs, several classic studies demonstrated
that such a cutoff does exist (\citealt{sch91,war94,ken95}). The Sloan Digital
Sky Survey (SDSS) has now found QSOs out to redshifts beyond 6, and analyses of
the space density (\citealt{fan01a,fan01b,fan01c}) provide the strongest
evidence to date of the drop in space density beyond $z = 3$. X-ray surveys now
appear to show that the X-ray QSO population exhibits a decline at high
redshifts similar to that found for optically-selected QSOs
\citep{has03,bar03,sil04}. But do all these observations indicate a real
diminution or -- at least at optical wavelengths -- could it be due to a dust
screen (\citealt{hei88,fal93})? It is here that radio-selected samples such as
the present one can provide a powerful check: if a significant diminution is
seen in the radio luminosity function, it cannot be the result of dust
obscuration. \cite{dun90} presented some evidence for just such a cutoff of the
radio luminosity function (RLF) for flat-spectrum (QSO-dominated) populations;
and an earlier analysis of a sub-sample from the present work \citep{sha96}
added confirmation. More recently \cite{vig03} defined a complete sample of 13
radio QSOs at $z \sim 4$, from which they concluded that the space density of
radio QSOs is a factor of $1.9 \pm 0.7$ smaller than that of similar QSOs at $z
\sim 2$. However, \cite{jar00} questioned these radio-QSO results, focussing on
the possible effects of spectral curvature.

A possible dust screen has serious implications for the interpretation of the
Hubble diagram for SN Ia supernovae. Assuming no obscuration, current results
from the SCP (Supernova Cosmology Project) collaboration \citep{kno03} and the
Hi-z team \citep{ton03} favour an $\Omega_m = 0.3,\ \Omega_\Lambda = 0.7$
universe. Two further related issues make delineation of the QSO epoch very
important: galaxy formation, and the reionization of the Universe.

\subsection{Galaxy formation}

The dramatic cosmic evolution of radio galaxies and QSOs stood as a curiosity
on its own for over 30 years since the birth of the idea \citep{ryl55}, clouded
as it was in the source-count controversy \citep{sch90}. It is relatively
recently that corresponding evolution has been delineated for the
star-formation rate \citep{lil96,mad96} and for galaxy evolution, particularly
blue galaxies \citep{ell99}. The correlation between star-formation rate and
AGN space density \citep{wal98} strongly suggested a physical connection
\citep{boy98}. Before the emergence of the Lilly-Madau plot of star-formation
history, it was recognized that the model of hierarchical galaxy development in
a Cold Dark Matter (CDM) Universe would result in a `quasar epoch'
\citep{hae93a,hae93b}. The issue of `quasar epoch' and `redshift cutoff' has
therefore assumed particular importance in consideration of galaxy formation in
low-density CDM universes. The very existence of any high-redshift QSOs sets
constraints on the epoch of formation of the first galaxies.
Haehnelt (1993) showed how the then-new COBE normalization \citep{smo92}
together with the QSO luminosity function at high redshifts as measured by
\cite{boy91}, provided substantial information on the initial fluctuation
spectrum and the matter mix. He found that the $z=4$ luminosity function
excluded an initial-spectrum index of $n \le 0.75$ or a Hot Dark Matter
fraction $\ge 25$ per~cent. Relevant to the current view of the
low-matter-density CDM Universe, he found that $\Omega_\Lambda \le 0.75$.
\cite{hae93b} developed a model for the evolution of the QSO population based
on the existence of $\sim$ 100 generations and linking the QSO phenomenon with
the hierarchical build-up of structure in the Universe. The evolution of host
objects is mirrored in the evolution of the mass of newly formed black holes;
only a moderate efficiency for formation of an average black hole is necessary
to model the luminosity function.
The model suggested that nearly all galaxies are likely to have passed through
a QSO phase. \cite{kau00} produced a more sophisticated model by incorporating
a simple scheme for the growth of supermassive black holes into the CDM
semi-analytic models that chart the formation and evolution of galaxies. In
addition to reproducing the observed relation between bulge luminosity and
black-hole mass in nearby galaxies \citep{mag98}, the model is able to mimic
the enormous increase in the QSO  population from redshift 0 to 2, a feature
that the Haehnelt-Rees model was able to describe only qualitatively.
Their conclusion: ``Our results strongly suggest that the evolution of
supermassive black holes, quasars and starburst galaxies is inextricably linked
to the hierarchical build-up of galaxies."

\subsection{Reionization}


The paradigm of hierarchical structure growth in a CDM universe has long
suggested that after the recombination epoch at $z \sim 1500$, the reionization
of the Universe took place at redshifts between 6 and 20 (e.g.
\citealt{gne97}). This reionization is predicted to be patchy and gradual
\citep{mir00}, although some models  indicate that it should happen quite
rapidly (e.g. \citealt{cen02,fan02}).
Two major observational advances support the `patchy and gradual' scenario.
Firstly, SDSS discovery of QSOs at redshifts of 6 or more
\citep{fan00,fan01a,bec01} has given a glimpse of what may be the end of the
epoch of reionization: the first complete \cite{gun65} trough has been observed
in the $z=6.29$ QSO SDSS~1030+0524 \citep{bec01,pen02} and a second has been
seen in $z=6.43$ QSO SDSS~J1148+5251 \citep{fan03}. There is disagreement as to
whether this marks the true end of reionization \citep{son02}; but the
suggestion is that it may be essentially complete by $z\sim6-7$. Secondly, the
detection of polarized anisotropies with the Wilkinson Microwave Anisotropy
Probe (WMAP) has resulted in a measurement of the optical depth $\tau \sim
0.17$ to Thompson scattering \citep{ben03,kog03}, implying a reionization
redshift of $17\pm5$. The CMB is sensitive to the onset of ionization, while
Gunn-Peterson troughs are sensitive to the late stages, the cleanup of
remaining HI atoms. Resolving the large uncertainties in these redshifts could
yet result in a rapid reionization scenario. Nevertheless several recent papers
(see e.g. \citealt{hai03}) address the complex and interacting suite of
physical mechanisms that may be at play during an extended `patchy and gradual'
reionization epoch $6 <z < 17$.

In either a fast or a gradual scenario, identifying the source of this
reionization as well as epoch is of vital importance for such interconnected
reasons as:

\begin{itemize}

\item The role of reionization in allowing protogalactic objects to cool into
stars,

\item The small-scale temperature fluctuations in the CMB and how these are
influenced by patchiness in the reionization, and

\item The epoch of the first generation of stars, or galaxies, or collapsed
black-hole systems.

\end{itemize}

It is most likely that the reionization is via {\it photoionization} by UV
radiation from stars or QSOs, rather than {\it collisional ionization} in e.g.
blast waves from the explosive deaths of Population III stars \citep{mad00}. It
may be possible to detect this reionization epoch directly as a step in the
background radiation at radio frequencies between 70 and 240 MHz (redshifted
21-cm HI) or in the infrared, 0.7 to 2.6 $\mu$m, from H recombination
\citep{sha99}. It may be possible with results from the Planck mission to
identify features in the CMB that identify what the predominant mechanisms are;
and it may be possible to detect the UV sources responsible for the ionizing
flux at $z \sim 10-20$ with the James Webb Space Telescope \citep{hai03}.

QSOs have long been prime candidates for this reionization. However the
apparent decline in space density (from the evidence summarized above and by
\citealt{mad99}), is inconsistent with this interpretation. \cite{mad00} showed
that in the face of this apparent diminution, UV luminosity functions of
Lyman-break galaxies (LBG) provide 4 times the estimated QSO contribution at $z
= 3$. It is now commonly accepted that such objects (or their progenitor
components) take on the mantle. The formation of short-lived massive stars in
such galaxies provides the UV photons \citep{hae01}, although QSOs may supply a
significant fraction of the UV background at lower redshifts.

Because the cooling time is long, the low-density IGM retains some memory of
when and how it was ionized. Several investigators have found a peak in
temperature of the IGM at $z \sim 3$ \citep{sch00,the02} close to the peak of
the `quasar epoch'. Moreover, observations of several QSOs at the wavelength of
HeII Ly$\alpha$ near $z=3$ suggest delayed reionization of He I, with the
process not yet complete by $z=3$ \citep{kri01}. The implication is that the
QSO ionizing photons coincident with the peak in activity both reionize HeI and
dump entropy into the IGM to raise its temperature.


In all of these aspects, it is clear that conclusions on ionizing flux from
QSOs are dependent on poorly determined high-power regions of luminosity
functions, on apparent cutoffs observed primarily in optically-selected
samples, and then only for the most luminous QSOs.

It is a primary purpose of this paper to determine the radio luminosity
function using the near-complete data of the present sample, and to examine the
evidence for a redshift cutoff. Before this, we discuss the populations
involved in the flat-spectrum sample by examining the $N(z)$ relation
(\S~\ref{nzsec}). Subsequent to the RLF determination in \S~\ref{rlfsec}, we
consider the issue of a redshift cutoff (\S~\ref{cutsec}), and the form of this
cutoff. In \S~\ref{evol} we construct an  overall picture of epoch dependence
of space density for radio-loud QSOs. We compare this with the parallel results
for QSOs selected at optical and X-ray wavelengths, and with the behaviour of
star-formation rate with epoch. The final section (\S~\ref{sum}) summarizes
results from this paper and our preceding two papers.


\section{The redshift distribution}
\label{nzsec}

For a sample of objects complete to some flux-density limit, the redshift
distribution, $N(z)$, gives preliminary information on the epoch of the
objects, and allows the most direct comparison with other samples. The redshift
distribution gives direct information on neither the luminosity function nor
its epoch dependence; however it provides essential data for use with other
data such as source counts to enable the construction of epoch-dependent
luminosity functions. There have been many versions of this. Most are a variant
on either the $V_{\rm max}$ method \citep{sch68} or the technique of defining
the {\it luminosity distribution} (\citealt{wal80a,wal83}), obtained when a
complete $N(z)$ is available at one flux-density level at least.

\begin{figure}[htb]

\vspace{10.5cm} \includegraphics{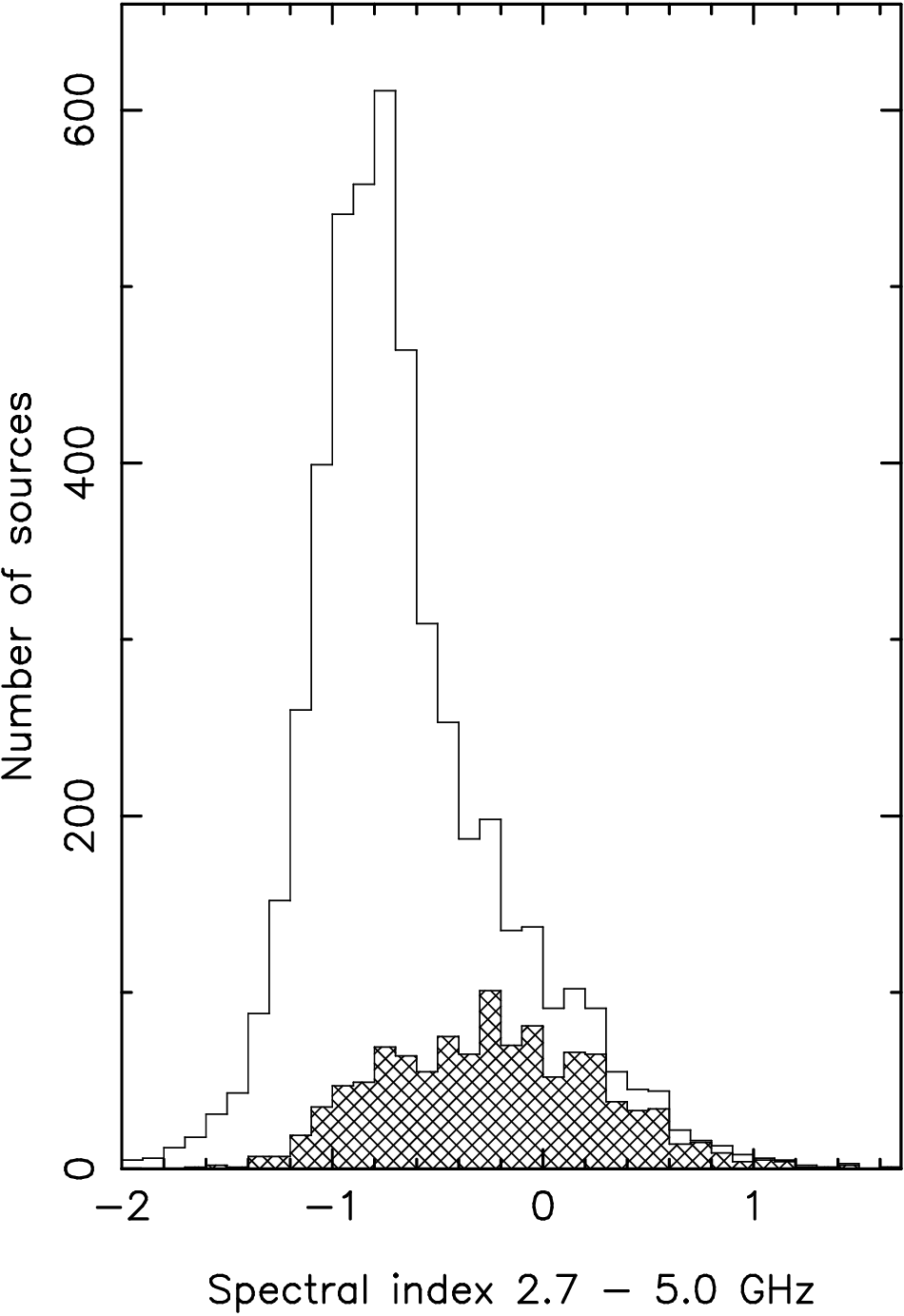}

\caption{The spectral index distribution for the compilation of sources from
Parkes surveys known as PKSCat90. The sub-sample selected has $S_{\rm 2.7\,GHz}
\ge 0.25$~Jy, in regions surveyed to a limiting flux density of 0.25~Jy or
less. The hatched area shows the sources identified as QSOs or BL\,Lac objects.
These beamed objects dominate the `flat-spectrum' region of the diagram.}

\label{spin}

\end{figure}

Such modelling processes now make use of statistical techniques to incorporate
data sets of varying completeness at many frequencies and flux-density levels.
The sample described here represents only one such data set, more complete than
most. \cite{dun90} carried out the most extensive such modelling. They took as
a starting point two populations, \emph{`flat-spectrum'} and
\emph{`steep-spectrum'} radio sources, now broadly considered in the light of
unified models as \emph{beamed} radio sources (radio-loud QSOs and BL\,Lac
objects) and their \emph{unbeamed} progenitors, or hosts (FRI and FRII radio
galaxies). All these objects are deemed to be powered by accretion-disk /
rotating black-hole systems from which a pair of opposing relativistic jets
feed double radio lobes. The single axis is collimated  during the feeding
process by rotation of the black-hole system. The beamed objects, QSOs and
BL\,Lacs, beamed because of relativistic ejections of components along axes
aligned with the line-of-sight, have radio structures dominated by
relativistically-boosted core emission. This core emission shows the effects of
synchrotron self-absorption and therefore has a flat or inverted radio
spectrum. The radio emission from powerful radio AGN whose axes are not aligned
with the line-of-sight is dominated by their steep-spectrum lobe emission, on
the large scales of 10's of Kpc up to 100's of Mpc. The dichotomy between
beamed and unbeamed objects as evidenced by their integrated radio spectra is
shown in Fig.~\ref{spin}.

The widely-used Dunlop-Peacock models of the luminosity functions may be simply
tested against the present data by means of the $N(z)$ distributions that they
predict.


We constructed a redshift distribution from the sample of Paper~1 as follows.
We selected all sources with $S_{\rm 2.7\,GHz} \ge 0.25$~Jy in regions for
which the 2.7-GHz flux-density limit was 0.25~Jy or less, and at declinations
$2.5\d \ge \delta \ge -40\d$. We refer to this as Sample~1 and the total area
it covers (Fig.~1 of Paper~1) is 2.676~sr. The source composition,
identification and redshift data for this sample are shown in Table~\ref{nz}.
Choice of the declination limit comes from both identification and
radio-spectral completeness; see \S 3.

\begin{table}[htb]
\begin{center}
\caption{Sample 1: an all-source sample selected from the source list of
Paper~1 in order to estimate the redshift distribution: $S_{\rm 2.7\,
GHz}\ge0.25$~Jy, $S_{\rm lim}\le0.25$~Jy, $2.5\d \ge \delta \ge -40\d$.}

\begin{tabular}{crrr}

\hline\hline
Ident & No Redshift & Redshift & Total \\

\hline
                QSO &  20 & 308 & 328\\
                BL  & 34    & 9 & 43\\
                G    &57    & 27 & 84\\
               Obsc    &  2    &  0 & 2\\
                e   &  1    &  0 & 1\\
\hline
                Totals &114    &344 & 458\\
\hline

\end{tabular}
\label{nz}
\end{center}
\end{table}

\begin{figure}[htb]

\vspace{8.5cm} \includegraphics{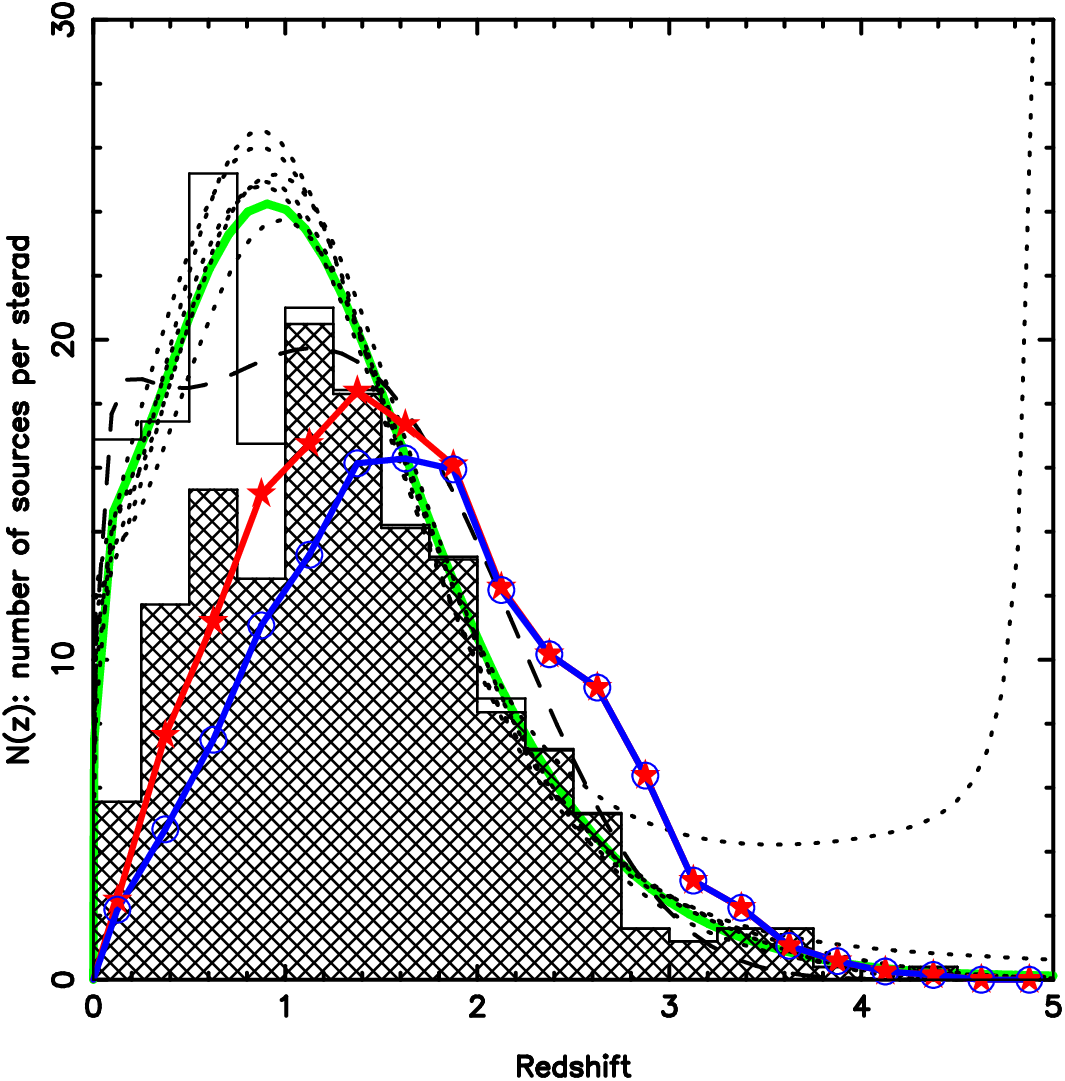}


\caption{The redshift distribution (histogram) for the sources of Sample~1
(Table~\ref{nz}). The hatched area shows the redshift distribution for beamed
objects alone, the QSOs + BL\,Lac objects, while the clear region represents
the galaxies. The 6 dotted lines show the appropriately-scaled distributions
predicted by the flat-spectrum ($\alpha \ge -0.5$) components of the
Dunlop-Peacock (1990) luminosity-function models, with the dashed line
distinguishing their pure-luminosity-evolution model. The solid line
represents the average of 6 of the models, omitting the model showing the very
steep rise to $z = 5$. The symbol + line systems show the predictions from the
dual-population models of Jackson \& Wall (1999), red representing
all beamed objects (QSOs + BL\,Lacs), blue for QSOs only.}

\label{nzfig1}

\end{figure}

 The entries in the identification column, Table~\ref{nz},
refer to (QSO)s, (BL)\,Lac objects, (G)alaxies, (Obsc)ured fields, and (e) not
identified for reasons discussed in Paper~1. As reasonable approximations, the
20 QSOs without measured redshifts were assumed to have the same redshift
distribution as those with redshifts; likewise the unmeasured redshifts of the
34 BL\,Lac objects were assumed to have the same distribution as those
measured. Such an approximation is inappropriate for the galaxies, however. A
crude Hubble diagram was plotted for the 27 galaxies with redshifts and a
simple polynomial was fitted to make rough estimates of the redshifts for the
remaining 57 galaxies based on their $B$ magnitudes. Finally the 3 Obsc and e
sources were assumed to have the same redshift distribution as the total
sample; the $N(z)$ obtained by adding the QSO, BL\,Lac and galaxy redshifts was
simply scaled by $(344+111+3)/(344+111)$ to obtain the final $N(z)$ of
Fig.~\ref{nzfig1}.


Dunlop \& Peacock derived luminosity functions for their two-population model,
flat-spectrum and steep-spectrum radio sources, representing the luminosity
functions as polynomials over the surface ($\rho, P_{\rm radio}, z$), and
obtaining coefficients by best-fitting to multi-frequency survey data including
source counts and redshifts. Different models resulted from different starting
points and factorizations of the epoch-dependent luminosity function. Their
division between flat-spectrum and steep-spectrum sub-populations took spectral
index $\alpha = -0.5$ as the dividing criterion. The predictions of redshift
distributions from the flat-spectrum portions of Dunlop-Peacock luminosity
functions are shown in Fig.~\ref{nzfig1}. In order to scale these to our
spectral-selection criterion, we used the spectral-index histogram of
Fig.~\ref{spin}; the ratio of objects with $\alpha \ge -0.4$ (our selection
criterion, Paper~1) to those with $\alpha \ge -0.5$ is 1060/1275 = 0.831.

In view of uncertainties in spectral index and of equating the flat-spectrum
population of Dunlop \& Peacock with compact radio sources, the overall
agreement is good. The form of the decline in $N(z)$ to higher redshifts is
impressively described by the Dunlop-Peacock models. Two models stand out in
Fig.~\ref{nzfig1}. One model with a space-density cutoff at $z = 5$ predicts a
redshift distribution greatly at variance with observations, showing a dominant
spike in the distribution at redshifts just below this cutoff. It has been left
out of the averaging process. The pure-luminosity-evolution model, shown as the
dashed line, is distinct in having a quicker rise and flatter maximum than the
others. These two features provide a better representation of the data in the
range $0<z<1.5$ than do the other models.

The good fit of the Dunlop-Peacock models to the total $N(z)$ distribution for
flat-spectrum objects does not imply a good description of the $N(z)$ for
beamed objects (hatched area, Fig.~\ref{nzfig1}) alone. The Dunlop-Peacock
models clearly rely on the presence of low-luminosity flat-spectrum galaxies
for the quality of overall fit; the `flat-spectrum' models describe the beamed
objects alone rather poorly.







Models considering populations in terms of beamed and host object were
developed by Jackson and Wall \citep{wal97,jac99}. The $N(z)$ predictions from
these models are shown in Fig.~\ref{nzfig1}. Agreement is reasonable;
normalization is correct, and the forms of the curves are similar. This
agreement is expected on the basis of the fit of the model to the 5-GHz source
count and the incorporation of a redshift cutoff in the model evolution. The
models over-predict objects at $z > 2$, due primarily to a lack of constraint
on the evolution of low-luminosity sources.

\section{The Radio Luminosity Function (RLF)}

\label{rlfsec}

Completeness of identifications enables the radio luminosity function to be
constructed in a straightforward way, using the $1/V_{obs}$ approach
\citep{sch68,fel76,avn80}. The contribution of each object to space density is
calculated as the reciprocal of the observable volume, the volume defined by
the redshift range(s) in which the object can be seen. Because the sample is
optically complete, only radio data (apart from the redshifts) are relevant in
defining this range.


An appropriate sub-sample for this calculation is that referred to as Sample~2
in Table~\ref{samtab}. Selected from the catalogue of Paper~1, it includes all
the QSO identifications with flux densities above survey limits and within the
declination range +2.5\d\ to $-$40\d. Defining $V_{obs}$ requires knowledge of
the radio spectrum both above and below the survey frequency. Above 2.7~GHz,
there are the 5.0-GHz data of the Parkes catalogues for \emph{all} sources in
the 2.7-GHz surveys, flux densities at 8.4~GHz for many of these sources
\citep{wri90b}, and about 40 8.87-GHz flux densities for some of the brighter
sources \citep{shi73}. Below 2.7~GHz, flux densities exist for most members of
Sample~2 at 365~MHz from the Texas survey \citep{dou96}, and at 1.4~GHz from
the NRAO VLA sky survey \citep{con98}.  The Texas survey covers the sky at
declinations down to $-$35.5\d\ and the NVSS down to $-$40\d. As a compromise
between sample size and spectral completeness, the sub-sample chosen for
definition of the RLF, Sample~2 of Table~\ref{samtab}, was therefore taken to
have a southern declination limit of $-$40\d. Most of the area surveyed at 2.7
GHz in this range has a completeness limit of $S_{\rm 2.7\,GHz} = 0.25$ Jy, but
some regions have limits of 0.10, 0.20 and 0.60 Jy; see Fig.~1 of Paper~1.

\begin{table}[htb]

\begin{center}
\caption{QSO samples for RLF and redshift-cutoff analyses.}
\begin{tabular}{lrrrr}
\hline\hline
Sub-sample & \multicolumn{2}{c}{Sample 2$^\dag$} & \multicolumn{2}{c}{Sample 3$^\ddag$}\\

\hline
&&&&\\
\ \ with measured z UKST ID & 342 &      &\ \  242 &     \\
\ \ with measured z CCD ID  &  13 &      &\ \   10 &     \\
total QSOs with measured z       &     &\ \  355  &     & 252 \\
&&&&\\
\ \ no measured z UKST ID   &  17 &      &\ \   11 &     \\
\ \ no measured z CCD ID    &   6 &      &\ \    4 &     \\
total QSOs with no measured z    &     &\ \   23  &     &  15 \\
&&&&\\
no ID (no measured z)       &   1 &   1  & \ \   1 &   1 \\
&&&&\\
Totals QSOs + non-ID         & 379 & 379  &\ \  268 & 268 \\

\hline
\end{tabular}
\label{samtab}
\end{center}
$^\dag S_{\rm 2.7\,GHz} \ge S_{\rm lim}$, $+2.5\d > \delta
> -40\d$, area 3.569 sr.\\
$^\ddag S_{\rm lim} =0.25$, $S_{\rm 2.7\,GHz} \ge S_{\rm lim}$, area 2.278 sr.
\end{table}

The steps to defining $V_{\rm obs}$ consist of (1) determining
$P_{2.7\rm~GHz}$, the luminosity of the radio source at 2.7~GHz (rest frame),
and (2) `moving' the source with its spectrum defined by the measured flux
densities, from $0<z<\infty$ to determine in which redshift range(s) it is
observable. It is observable at a given  redshift if (a) its flux density
exceeds the survey limit $S_{2.7} = 0.25$~Jy and (b) its redshifted spectrum
over the observer's range 2.7 to 5.0~GHz has a spectral index $\ge -0.4$. We
interpolated between measured spectral points in the log~S$_\nu$ - log~$\nu$
plane. Despite the relatively sparse sampling in this plane, combined
luminosity and spectral effects of `moving' the source are complex, sometimes
resulting in a source having two regions of observable volume defined by four
redshifts. (These effects are discussed further in the following section.) In
calculating the RLF, the contribution of each source is then
\[\sum_i 1/(V^i_{\rm max} - V^i_{\rm min})\] where $i$ is usually unity but is
sometimes two. Throughout the analyses we have used the geometry $H_0 = 70$ km
s$^{-1}$ Mpc$^{-1}$, $\Omega_{\rm tot} = 1.0$, $\Omega_{\rm m} = 0.3$,
$\Omega_\Lambda = 0.7$.

\begin{table*}[htb]
\begin{center}
\caption{The radio luminosity function $\rho$, in units of log(Mpc$^{-3}$) per
$\Delta z = 0.5$ per $\Delta ($log$P) = 0.4$, as derived from Sample~2,
Table~\ref{samtab}. $N$ is the number observed per bin and $\overline{z}$ the
mean redshift of the sources in the bin.}
\begin{tabular}{crrrrrrrrrrrrrrr}

\hline\hline

 Power & \multicolumn{3}{c}{$0<z<0.5$} &
\multicolumn{3}{c}{$0.5<z<1.0$} & \multicolumn{3}{c}{$1.0<z<2.0$}
& \multicolumn{3}{c}{$2.0<z<3.0$} & \multicolumn{3}{c}{$3.0<z<5.0$}\\

log(P/
WHz$^{-1}$sr$^{-1})$&log$\rho$&$N$&$\bar{z}$&log$\rho$&$N$&$\bar{z}$&log$\rho$&$N$
&$\bar{z}$&log$\rho$&$N$&$\bar{z}$&log$\rho$&$N$&$\bar{z}$\\

\hline

24.8 &  $-$9.02 & 4 &0.33 & $-$10.22 & 1 &0.58 &---    &0 &--- &--- &0 &---
&---
&0 &---\\
25.2 &  $-$8.60 &15 &0.36 &  $-$9.13 & 6 &0.62 &---    &0  &--- &--- &0 &---
&---
&0  &---\\
25.6 &  $-$9.54 & 2 &0.32 &  $-$8.91 &29 &0.73 & $-$10.40 & 5 &1.11 & $-$11.43
& 1
&2.12 &---   &0  &---\\
26.0 & $-$10.27 & 1 &0.36 &  $-$9.26 &19 &0.77 & $-$9.57 &45 &1.32 &---    &0
&---
&---   &0  &---\\
26.4 & $-$11.11 & 1 &0.16 &  $-$9.85 &15 &0.78 & $-$9.33 &91 &1.47 & $-$10.34
&19 &2.24
& $-$12.13 & 1 &3.45\\
26.8 &---     &0  &--- & $-$10.35 & 5 &0.77 & $-$10.05 &29 &1.69 & $-$10.05 &34
&2.48
& $-$11.02 &12 &3.47\\
27.2 &---     &0  &--- &---     &0  &--- & $-$11.00 & 6 &1.50 & $-$10.98 & 8
&2.33
& $-$11.82 & 2 &3.45\\
27.6 &---     &0  &--- &---     &0  &--- &---    &0  &--- & $-$12.19 & 1 &2.56
& $-$12.32 & 1 &3.57\\

\hline
\end{tabular}
\end{center}
\label{rlf_tab}
\end{table*}

Following these precepts, the radio luminosity functions calculated for
rest-frame powers at 2.7~GHz for 5 redshift ranges are given in
Table~\ref{rlf_tab}.


\begin{figure*}[htb]

\vspace{8.0cm}



\includegraphics{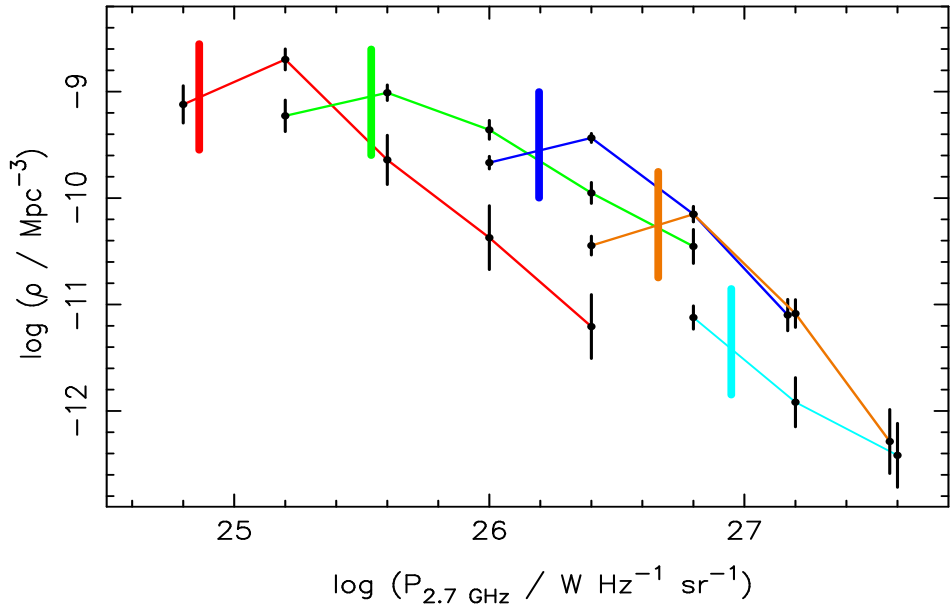}

\caption{The radio luminosity function ($H_0 = 70$ km s$^{-1}$ Mpc$^{-1}$,
$\Omega_{\rm tot} = 1.0,\ \Omega_{\rm m} = 0.3,\ \Omega_\Lambda = 0.7$) for the
QSOs of the Parkes 0.25-Jy flat-spectrum sample. The RLF was computed in
redshift ranges $0 - 0.5$ (red), $0.5 - 1.0$ (green), $1.0 - 2.0$ (blue), $2.0
- 3.0$ (orange) and $3.0 - 5.0$ (light blue). Vertical bars show the limits of
completeness in power for each redshift range, as described in the text. Error
bars are $1/\sqrt{N}$ with $N$ from Table~\ref{rlf_tab}. Sky area and
redshift-measurement completeness have been considered in order to plot true
space densities per $\Delta z = 0.5$, $\Delta$(log$P$)$=0.4$.}

\label{rlf_dif}

\end{figure*}
\begin{figure}[htb]

\vspace{7.5cm} \includegraphics{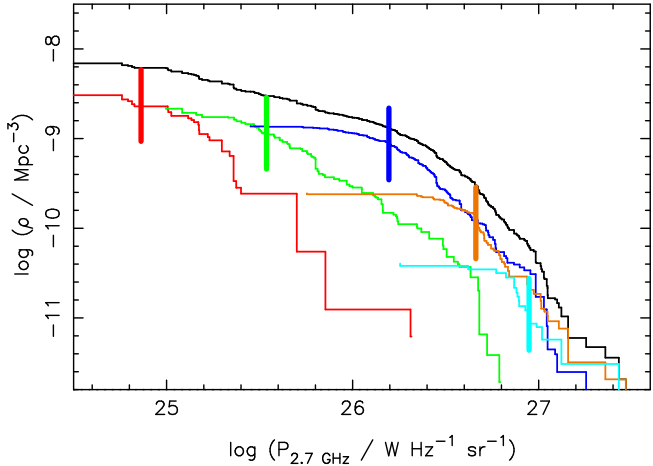}


\caption{The integral radio luminosity function  for the QSOs of the Parkes
0.25-Jy flat-spectrum sample, computed in the five redshift ranges of
Fig.~\ref{rlf_dif}: $0 - 0.5$ (red), $0.5 - 1.0$ (green), $1.0 - 2.0$ (blue),
$2.0 - 3.0$ (orange) and $3.0 - 5.0$ (light blue). Vertical lines again
indicate limits of completeness for each redshift range, due to spectral-index
spread. The upper curve is the total integrated luminosity function, complete
for all powers only at the very highest luminosities.}

\label{rlf_int}

\end{figure}

These data are displayed in Fig.~\ref{rlf_dif}. First impressions are that the
curves slide sideways, suggesting simple luminosity evolution, as deduced from
similar behaviour in redshift shells for optical luminosity functions
\citep{boy88}. However, the transition from the lowest redshift range ($z <
0.5$) to the next redshift range ($0.5 < z < 1.0$) is not described by a
lateral shift. It is possible that the lower bin is contaminated by unbeamed
objects such as Seyfert galaxies and elliptical-galaxy cores; such objects may
have entirely different central engines and different luminosity functions as a
result. The data of Fig.~\ref{rlf_dif} shown in integral form
(Fig.~\ref{rlf_int}) suggest that the RLF changes in form right out to the $1 <
z < 2$ shell, and it is improbable that contamination by unbeamed objects
persists beyond $z > 0.5$. A closer investigation by both morphology and
spectrum is needed to determine if removal of unbeamed objects could `save'
luminosity evolution. However, it is probably beyond saving. For example, from
QSOs discovered in the SDSS survey, \cite{fan01b} noted that the high-power end
of the QSO luminosity function appears flatter than that at lower redshifts.

The impression of luminosity evolution may be misleading in any case. Spectral
spread limits the upper power bound of completeness for the RLF in each
redshift band. At the maximum redshift of the band, radio sources of the
steepest spectra fall below the survey flux-density limit first; the power
limit is determined simply from

\[
P_{lim} = S_{lim} \times D^2 \times (1 + z)^{(1 - \alpha_{max})}
\]

\noindent where $D$ is the `luminosity distance', and $\alpha_{max}$ is the
minimum (low-frequency) spectral index, i.e. that effective index corresponding
to the source with `steepest' radio spectrum in the particular redshift range.
At lower powers within the bin, the RLF will be incomplete for such objects,
but will remain complete for objects of flatter spectra. (The limit is well
defined for our sample; we selected objects of $\alpha \ge -0.4$, i.e. the
spectral limit was imposed on the `steep' side, with of course no limit as to
how `flat' or `inverted' the spectra might be.) This limit may cause RLFs of
similar slopes to appear to have a knee at similar space densities, mimicking
luminosity evolution. In previous discussions of space densities it is not
clear that this limit plus spectral spread have been considered; several such
studies appear to ascribe a single canonical spectrum to every QSO.

One regrettable result of this power limit is that tracing the space densities
in the higher-redshift ranges down to low powers is not possible. Composite
RLFs (galaxies plus QSOs) extending over many decades show relatively few QSOs
at low redshifts, where the RLFs are dominated by low-luminosity (mostly
star-forming) radio galaxies \citep{sad02}. The RLFs of QSOs at high redshifts
must therefore flatten and drop drastically towards the lower powers. The
dual-population models of \cite{jac99} demonstrate such behaviour. From the
present data, the limit-lines show that the only conclusion to be drawn is that
the RLFs  may reduce in slope towards the lower powers. In \S \ref{evol} we
show how a different approach can yield some information throughout the range
of redshifts occupied by the present sample.

\begin{figure}[htb]

\vspace{7.7cm} \includegraphics{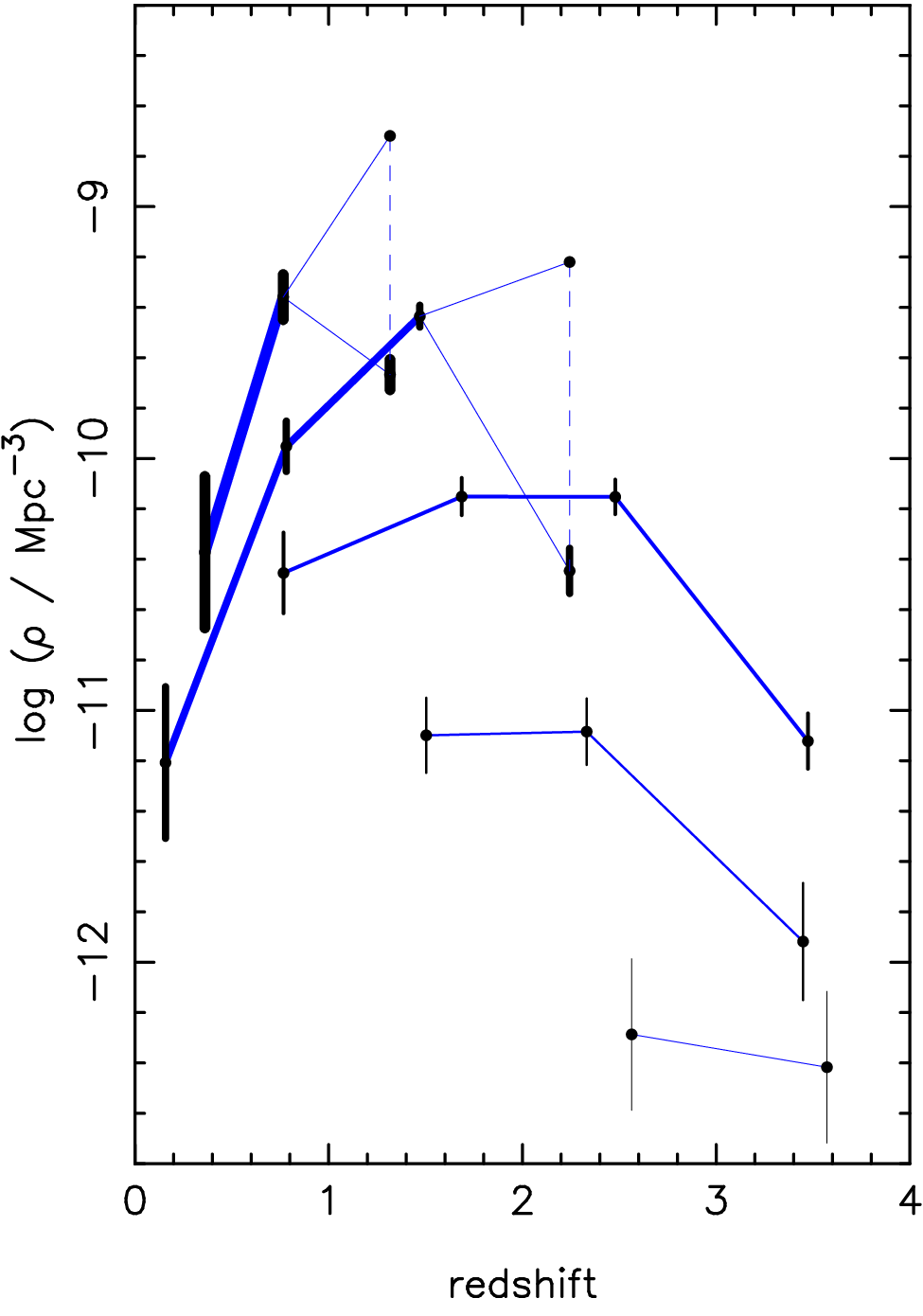}


\caption{Space densities as a function of redshift for 5 power ranges,
log$P_{2.7} = 25.8 - 26.2$, $26.2 - 26.6$, $26.6 - 27.0$, $27.0 - 27.4$, and
$27.4 - 27.8$, denoted by decreasing line weight. Abscissa values are the mean
redshifts in each element of ($\Delta P_{2.7}$, $\Delta z$). For each of the
two lowest power ranges, the final point with error bar represents incomplete
data, as these points fall at powers below the spectral cutoffs shown in
Fig.~\ref{rlf_dif}. Upper limits, represented by the single dots, were obtained
by extrapolating the RLF to this power from higher powers. The actual values
therefore lie somewhere along the two dashed lines.}

\label{rlf_z}

\end{figure}

A third presentation of the RLF data is given in Fig.~\ref{rlf_z}, in which
space densities are plotted as a function of redshift for 5 ranges of intrinsic
power. The initial dramatic increase in space density with redshift is evident,
with densities in the redshift range $1.0 - 2.0$ some two orders of magnitude
above those for objects at redshifts $< 0.5$. Small numbers at the highest
redshifts (see Table~\ref{rlf_tab}) and the completeness limits at the lower
redshifts constrain the redshift range observable for each luminosity. In
particular it is not possible to judge whether the maximum space density is a
function of radio luminosity. The curves overlap adequately to show
self-consistency, and to demonstrate the increase in space density from small
redshifts to $z \sim 1.5$. Beyond this redshift, the space densities for each
power range decline, although statistical uncertainties are substantial.
Fig.~\ref{rlf_dif} also indicates such a decline; these data therefore suggest
a redshift cutoff, at some level of significance.


\section{The Redshift Cutoff}

\label{cutsec}

Our preliminary analysis \citep{sha96} indicated a decrease in radio-QSO space
density beyond $z = 3$. Using a well-defined sub-sample from the present study,
Shaver et al. considered the space density of QSOs with $P_{\rm 2.7\,GHz} \ge
1.1 \times 10^{27}$ W Hz$^{-1}$ sr$^{-1}$. On the basis of uniform space
density, the 25 such radio QSOs seen at $z \le 4$ indicate that 15 similar
objects would be expected in the range $5 \le z \le 7$. None was found. From
Poisson statistics, the difference is significant at the 99.9\% level.

\begin{figure}[htb]

\vspace{7.0cm}

\includegraphics{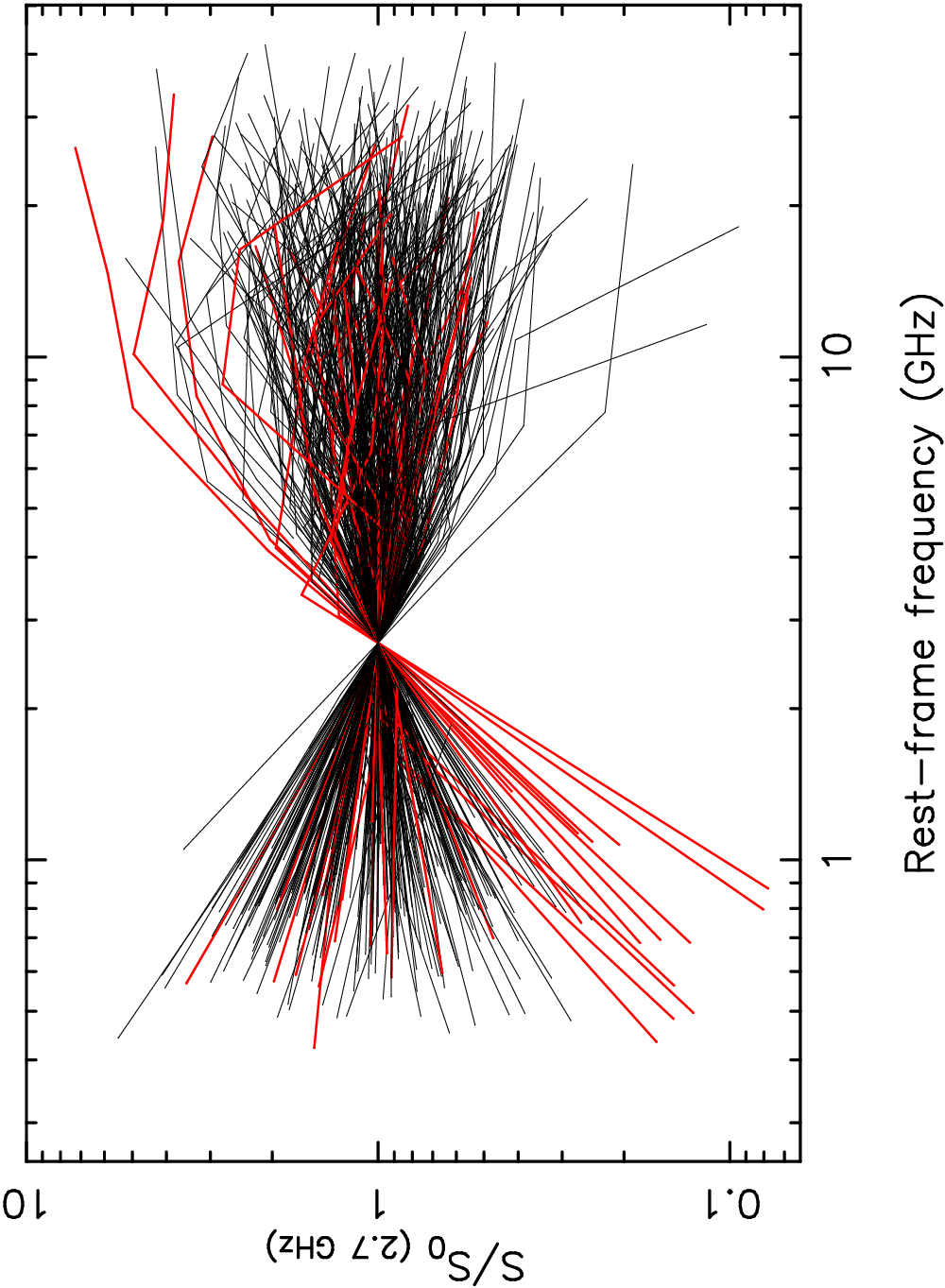}

\caption{The radio spectra of all sources in Sample~2, Table~\ref{samtab}, in
their rest frame. Data are at observing frequencies of 0.365, 1.4, 2.7, 5.0,
8.40 and 8.87 GHz, and flux densities are normalized by $S_{0}$, the
interpolated rest-frame flux density at 2.7~GHz. The red lines represent the
brightest sources, those with $S_{\rm 2.7\,GHz}\ge 2.0$ Jy.}

\label{mother}

\end{figure}

This preliminary study drew attention to a possible difficulty in the analysis
due to the curved nature of some of the radio spectra. \cite{jar00} examined
this in some detail, pointing out the apparently curved nature of many of the
radio spectra involved, and indicating how such an effect, a steepening to the
high frequencies in particular, might reduce or remove the significance of an
apparent redshift cutoff. Their model-dependent analysis used only the
highest-power objects and indicated that the apparent cutoff on the basis of
such objects might have a significance level as low as that corresponding to
$2\sigma$. They suggested that establishing the reality of the cutoff for such
objects to a high level of significance might be difficult even with all-sky
samples. However, Fig.~\ref{mother} shows that there is no clear majority of
sources with spectra steepening to the higher frequencies. Moreover, we show
below that the spectral data in the literature are misleading in terms of the
proportion of sources showing spectral steepening to the higher frequencies.

Subsequently we have considered alternative methods to study space density and
redshift distribution, methods to utilize the entire sample which can
demonstrate simple attributes of the space-distribution without recourse to
modelling the luminosity function or its epoch dependence.

\subsection{The Power-Volume plane; using the whole sample}

\begin{figure*}[htb]

\vspace{12cm} \includegraphics{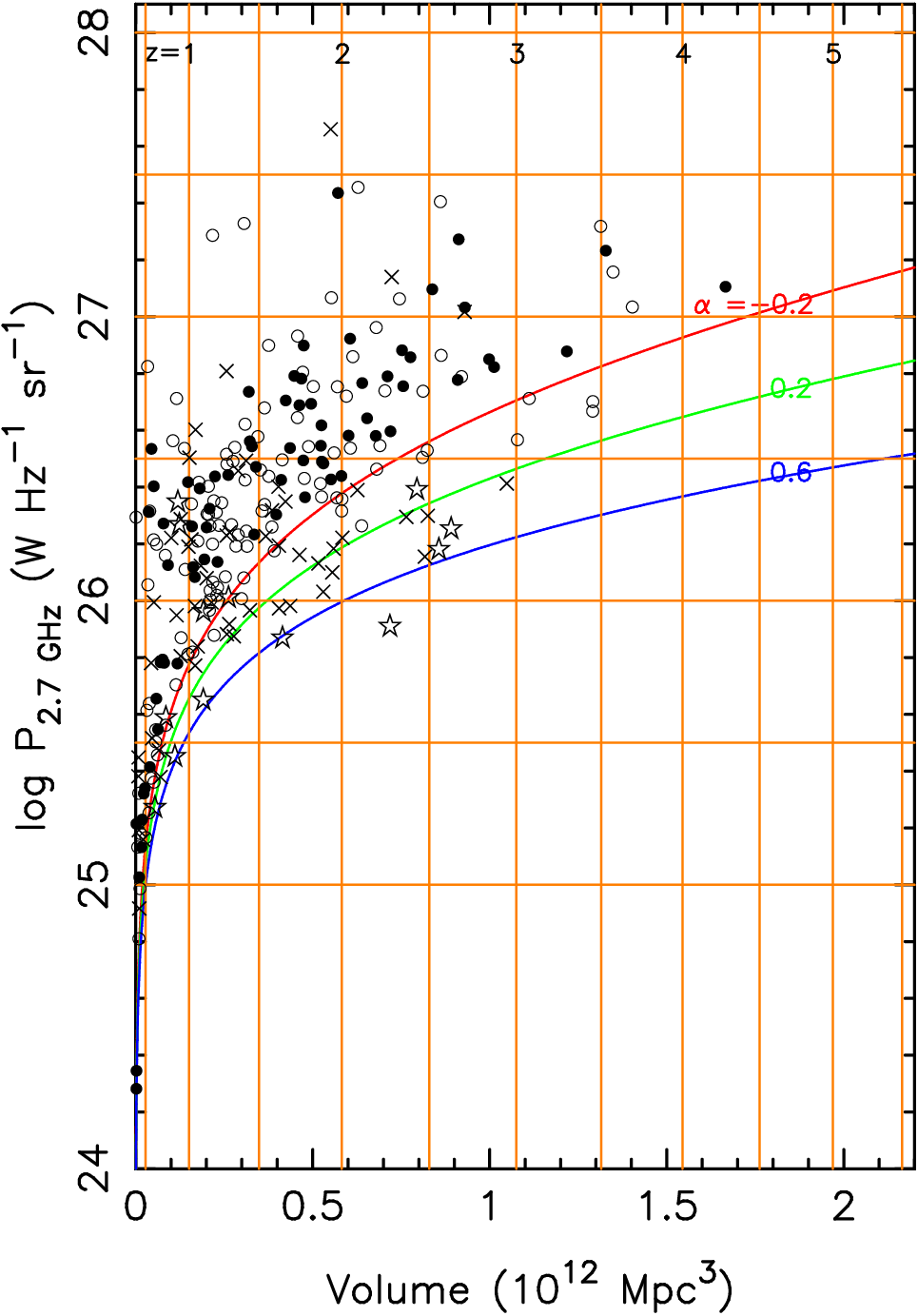} \includegraphics{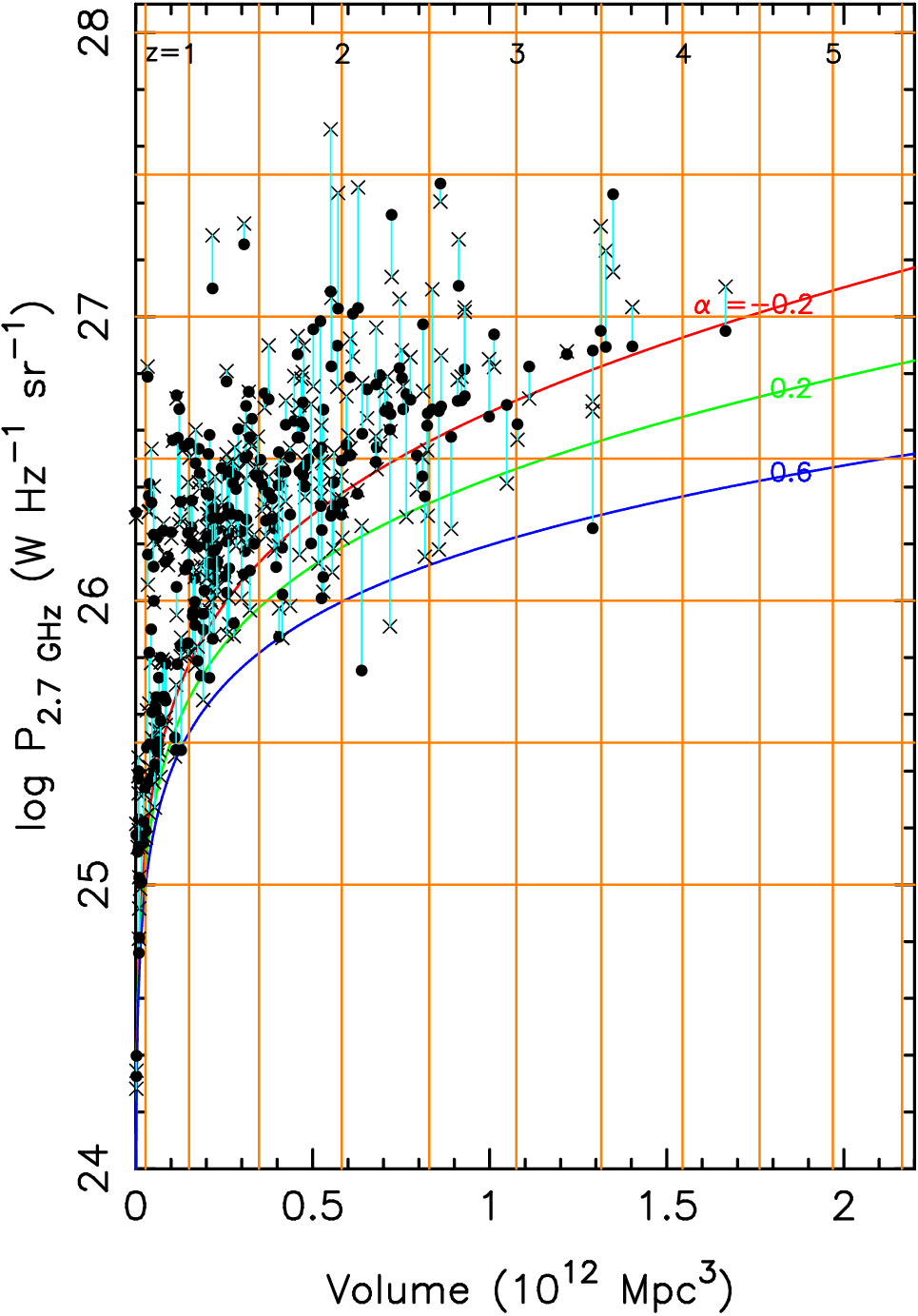}


\caption{The Luminosity - Volume plane for the 252 QSOs with measured
redshifts, in survey areas with completeness limit $S_{\rm 2.7\,GHz} = 0.25$~Jy
(Sample~3, Table~\ref{samtab}). Vertical grid lines indicate redshifts, as
marked along the top border. The curved lines indicate survey completeness
limits at 0.25~Jy for three spectral indices. Left: the sources plotted with
symbols to indicate different spectral indices $\alpha^{2.7}_{5.0 {\rm GHz}}$:
solid dots for $-0.4 <\alpha < -0.2$; open circles for $-0.2 < \alpha < +0.2$,
crosses for $+0.2 < \alpha < +0.6$, and stars for the extreme spectral
inversions $\alpha >+0.6$. Rest-frame luminosities ($P_{2.7}$) for this plot
were calculated assuming a power-law spectrum given by this index. Note that
the dots lie above the $\alpha = -0.2$ limit line; the open circles above the
$+0.2$ line; the crosses above the $+0.6$ line; and the stars scatter to below
the $+0.6$ line. Right: The effects of considering spectral data at frequencies
below the Parkes survey frequency (2.7~GHz). The values of $P_{2.7}$ are
substantially changed. Crosses indicate the original positions as in the left
panel, while dots show the positions revised with improved estimates due to
incorporation of lower-frequency data.}

\label{p-v}

\end{figure*}

Fig.~\ref{p-v} (left) shows sources of Sample~3 (Table~\ref{samtab}) in a plot
of radio luminosity vs. co-moving volume. We need this new sample for such a
plot. Recall that Sample~1 (Table~\ref{nz}) included all sources, not just
QSOs, while Sample~2 (Table~\ref{samtab}), although confined to QSOs, was drawn
from regions of the survey with different completeness limits. In order for a
plot of luminosity vs. $z$ (or equivalently, co-moving volume) to be
interpreted, the sample must have a single survey limit. Sample~3 is therefore
composed of all QSOs from our data table of Paper~1 with survey completeness
limit at exactly $S_{\rm 2.7~GHz} = 0.25$~Jy (Fig.~1, Paper~1), and again at
declinations above $-40\d$ for reasons of radio-spectral completeness.
Fig.~\ref{p-v} shows lines of survey completeness corresponding to 0.25~Jy for
three different radio spectral indices.

The plot with co-moving volume on the abscissa rather than redshift gives
direct indication of space density. There is an apparent diminution in the
density of points at redshifts above $\sim 2.5$. The question is whether this
is real and significant. In what follows we test the null hypothesis that the
space density of QSOs at high redshifts remains constant and equal to that at
$1 < z < 3$.

Redshift information is not complete for Sample~3; in order to make comparison
with prediction we must estimate the number of possible objects at $z
> 3$. Table~\ref{samtab} presents the summary. The key element is the
sub-sample of 16 objects in the sample of 268 for which redshifts are not
available.

The redshift distribution for QSOs is known to be a function of both apparent
magnitude and flux density, albeit with huge scatter and only a gentle
dependence in each case. Thus in order to estimate redshift proportions for the
objects without such data, we treated the identifications made on UKST plates
and those from the (deeper) CCD observations separately.

Consider the 11 QSOs without redshifts and identified from UKST plates. Of the
242 objects identified on UKST plates and with measured redshifts, 8 have $z >
3$. There is no bias in the redshift measurements or lack of; and thus we
expect $8/242 \times 11 = 0.36$ of the 11 objects to have $z > 3$. The
remaining 5 objects may be treated equally; the single unidentified source in
the sample (PKS 0225-065) escaped the CCD identification programme by being
de-identified later on the basis of an improved radio position. Had it been
included we can be confident that an identification would have been obtained,
as it was in each of the 87 cases we tried. For these 5 objects, then, we use
the CCD-identified QSOs with redshifts, totalling 10 in the sample, for which
two redshifts exceeded 3. We thus anticipate $2/10 \times 5 = 1.0$ of the 5
objects will have $z > 3$. The number of objects in the sample with measured
$z>3$ is 10. Thus the number with which to compare predictions for $z>3$ is
$10$(observed)$ + 1.4$(estimated)$ = 11.4$. The principal point is that
redshift incompleteness does not impede our analysis.

A simple analysis may be carried through on the basis of Fig.~\ref{p-v}. If we
consider QSOs in specific narrow bins of luminosity and survey limit imposed by
spectral index and survey flux limit $S_{\rm lim}$, then such horizontal
stripes in Fig.~\ref{p-v} intersecting the curved survey-cutoff lines provide
an area in the figure in which QSOs can be seen by the survey. On the null
hypothesis, no redshift diminution, if we now split this area into a region
with $z < 3$ and a region with $z > 3$, we can use the surface density of QSOs
in the low-redshift area to form an expectation value for the higher redshift
area. We chose the prediction region to be $1<z<3$ to coincide roughly with the
plateau of the `quasar epoch', and we selected the high-redshift region to run
out to $z=8$, the approximate limit to which we could hope to see QSOs given
our survey limits and the known range of luminosity and spectral index.

This process described above can be refined by reducing the stripes of radio
power to zero width; each source then becomes a predictor, provided of course
that the survey limit allows it to be seen beyond a redshift of 3.
Table~\ref{results} presents results of this analysis under the sub-heading
`single survey cutoff'. The immediate result is the apparent one: a prediction
of significantly more QSOs at $z>3$ than the 11.4 `seen'.

\begin{table*}[htb]
\begin{center}
\caption{Predicted numbers of QSOs, $3<z<8$ }
\begin{tabular}{ccccc}
\hline\hline
Sample & Observed at $z>3$ & Analysis type& Contributors & Predicted$^\dag~3<z<8$ \\

\hline
&&&&\\
\multicolumn{3}{l}{Single survey cutoff:} &    &      \\
3 &11.4 & single cutoff, $\alpha = +0.2$ & 73(86) & 64.7(66.7) \\
3 &11.4 & single cutoff, $\alpha = \ \ 0.0$ & 49(57) & 40.3(40.4) \\
3 &11.4 & single cutoff, $\alpha = -0.2$ & 39(43) & 25.0(25.1) \\
&&&&\\
\multicolumn{3}{l}{Source-by-source analysis, `Single-source survey':}  &    &      \\
2 &17.8 &  $S_{2.7}, S_{5.0}$ & 71(89) & 51.5(53.5) \\
2 &17.8 &  $S_{0.365}, S_{1.40}, S_{2.7}, S_{5.0}$ & 67(89) & 48.5(54.0) \\
2 &17.8 &  $S_{0.365}, S_{1.40}, S_{2.7}, S_{5.0}, S_{8.40}$ & 53(65) & 28.8(30.1) \\
2 &17.8 &  $S_{0.365}, S_{1.40}, S_{2.7}, S_{5.0}, S_{8.87}$ or $S_{8.40}$& 57(70) & 38.6(38.6) \\
2 &17.8 &  $S_{0.365}, S_{1.40}, S_{2.7}, S_{5.0}, S_{8.87}$ & 70(88) & 50.5(52.9) \\
&&&&\\
\hline
\end{tabular}
\label{results}
\end{center}
$^\dag$calculated in the geometry adopted for this paper; bracketed numbers are
for $\Omega_{\rm m} = \Omega_{\rm tot} = 1.0$, the Einstein - de Sitter
Universe
\end{table*}

The results reveal a fundamental flaw of this analysis, namely what limit line
to adopt, corresponding to which spectral index. It is apparent from
Fig.~\ref{p-v} that adopting $\alpha = -0.2$ is extreme; but even confining the
analysis to narrow bands of spectral index does not define where within that
band the survey cutoff or completeness line should be placed. The analysis at
this point appears to confirm what the eye sees in Fig.~\ref{p-v}, but shows
that taking the figure at face value is dangerous. Moreover here we have used
the $2.7-5.0$~GHz spectral index, characterizing each spectrum as a single
power law; spectral curvature or indeed any complexity of radio spectrum has
not been considered. (For low-frequency surveys, the spectral-index issue is
not so important, because most sources detected in them have power-law spectra
characterized by an index close to $-0.75$. In corresponding $P-z$ or $P-V$
planes, most sources from low-frequency surveys cluster closely along or just
to the left of the single limit line given by this spectral index.)

The analysis of \cite{sha96} attempted to circumvent the difficulties by
sticking to powers so high that the observational cutoff, the survey
completeness limit, did not come in to play. In doing so, the available
sub-sample becomes small and the statistical uncertainties are inevitably
larger.

These difficulties suggest the following refinement.

\subsection{Source-by-source analysis: the `Single-Source Survey'}

There is no need to stick to a single survey-limit line in the $P-V$ plane.
Each source can be considered alone, conceptually the result of a survey which
found it as a single source. For each such `single-source survey', a limit line
may be drawn in the plane {\it peculiar to that object and incorporating all
its radio-spectral information}. The prediction of this object for sources at
redshifts above 3 may then be added to the predictions from all `single-source
surveys' to derive a prediction total. In effect this is using the $V_{\rm
max}$ method to predict the number of objects in volumes at higher redshift on
the hypothesis that space density is uniform; it is doing so using the spectral
properties of each source individually.

A further advantage in such a process is that there is no longer a need to
stick to a sample defined by a single flux-density limit. To improve
statistical weight, all zones of the survey can be used, no matter what the
flux-density limit, provided of course that the value of the 2.7-GHz flux
density is greater than or equal to the completeness limit for the area in
which it was detected. (Sources for which this is not the case were marked in
the data-table of Paper~1.)
Each source in this analysis contributes a predicted number of sources given by
the ratio of its accessible co-moving volume in the redshift range $3<z<8$ to
that in the range $1<z<3$. The sum of all such predicted sources, based on all
sources observed in the redshift range $1 < z < 3$, gives us the total number
of $3<z<8$ sources expected in the survey for a constant comoving space
density.

A sample appropriate to this analysis is Sample~2 of Table~\ref{samtab}, giving
a total of 379 radio QSOs, 355 with measured redshifts. From an analysis
analogous to that carried out for Sample~3, we estimate that complete
identification and redshift data would add 1.8 sources to the 16 members of
this sample observed to have $z > 3$.

As a basic analysis of this type, when individual limits are applied as
described, using the $\alpha^{5.0}_{\rm 2.7\,GHz}$ spectral index appropriate
to each source, a prediction of 51.5 sources in the redshift range $3 < z < 8$
is obtained (Table~\ref{results}), c.f. the 17.8 sources `observed'. However a
particularly important feature of the approach is that it enables incorporation
of \emph{all} radio spectral data. Other investigators (in particular
\citealt{jar00}) have emphasized how important the form of the spectrum is. As
mentioned in the previous section, there are several effects on space density
analysis and these are illustrated in Figs.~\ref{p-v} and \ref{p-vdemo}:

\begin{enumerate}

\item
If the spectrum decreases to low frequencies, perhaps having a
   low-frequency cutoff due to synchrotron or free-free absorption, the
   effective power at the rest-frame survey frequency is reduced. The
   result may be a substantially lower position in the $P-V$ plane. This
reduces the `headroom' the object has to predict significant
   contribution at higher redshift; and it may remove it from the
   `prediction contributors' list entirely. Significant steepening to the lower
   frequencies of course has the opposite effect, raising its rest-frame
   power, its position in the $P-V$ plane, and increasing its prediction.

We have incorporated spectral data at 1.4 GHz from the NVSS survey
\citep{con98} and at 0.365~GHz from the Texas Survey \citep{dou96} to define
the low-frequency spectra for the majority of sources in Samples~2 and~3. The
results may be seen in Fig.~\ref{p-v}. The extreme-power objects have convex
spectra and drop down into the pack; but in addition, a number of less-luminous
objects rise by virtue of steep low-frequency spectra. When the prediction is
made incorporating the low-frequency data (Table~\ref{results}), these effects
approximately cancel out and the result differs little from the previous
estimate: 48.5 objects should be seen in the sample at $3 < z < 8$, c.f. the
`observed' number of 17.8.

\item
A more substantial difference is produced by the incorporation of spectral data
at frequencies higher than 5.0 GHz. \cite{sha96} pointed out that
high-frequency flux densities measured quasi-simultaneously by \cite{gea94}
indicated little spectral steepening; and certainly not enough to exclude
very-high-redshift objects. However the \cite{gea94} sample may not be
representative. For PKS sources a set of flux densities at 8.4~GHz was measured
by \cite{wri90b}, including a large fraction of the sources in both Sample~2
and Sample~3. These data suggest that spectral steepening is more common than
indicated by the \cite{gea94} measurements, although as Fig.~\ref{mother}
shows, it is not a feature of the majority of sources.

Spectral steepening beyond 5~GHz in the observer frame has two effects. First
it moves the cutoff line upward (Fig.~\ref{p-v}) so that the object in question
drops from the sample at relatively lower redshifts. Secondly when `moving' the
object to some redshift above the observed redshift, the spectrum becomes
steeper than the apparent 2.7 -- 5.0-GHz spectral limit of $-0.4$ used to
define the original sample; `flat-spectrum' objects whose spectra steepen
beyond 5.0~GHz (observer frame) become undetected as such at higher redshifts.

To consider the first of these two effects, the cutoff (survey limit) line for
each source was calculated using each `segment' of the spectrum as redshift is
changed. This simple interpolation in the log~$S_\nu$ - log~$\nu$ plane results
in segmented cutoff lines for each object in the $P-V$ plane as shown in
Fig.~\ref{p-vdemo}.
\begin{figure*}[htb]

\vspace{12cm} \includegraphics{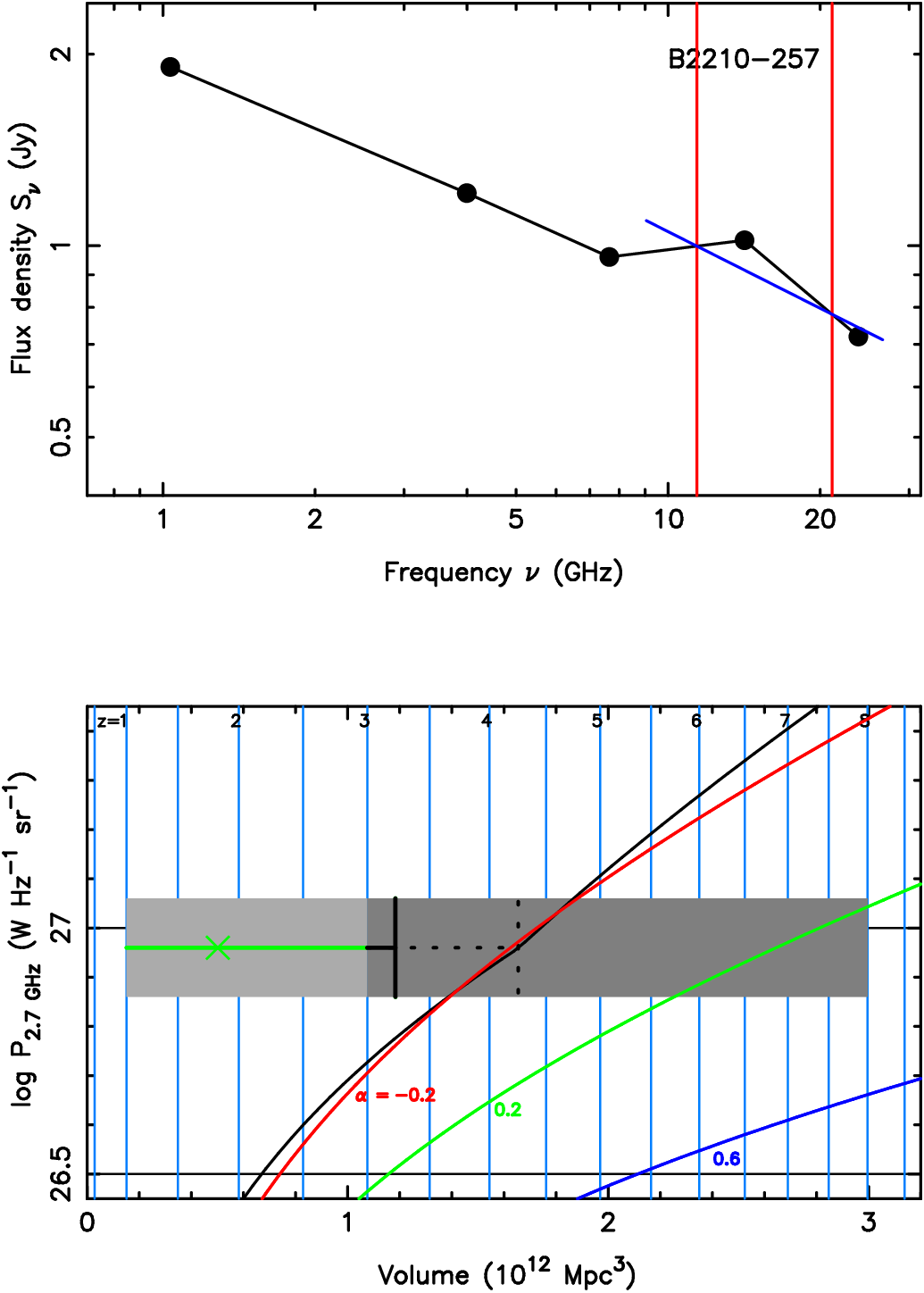} \includegraphics{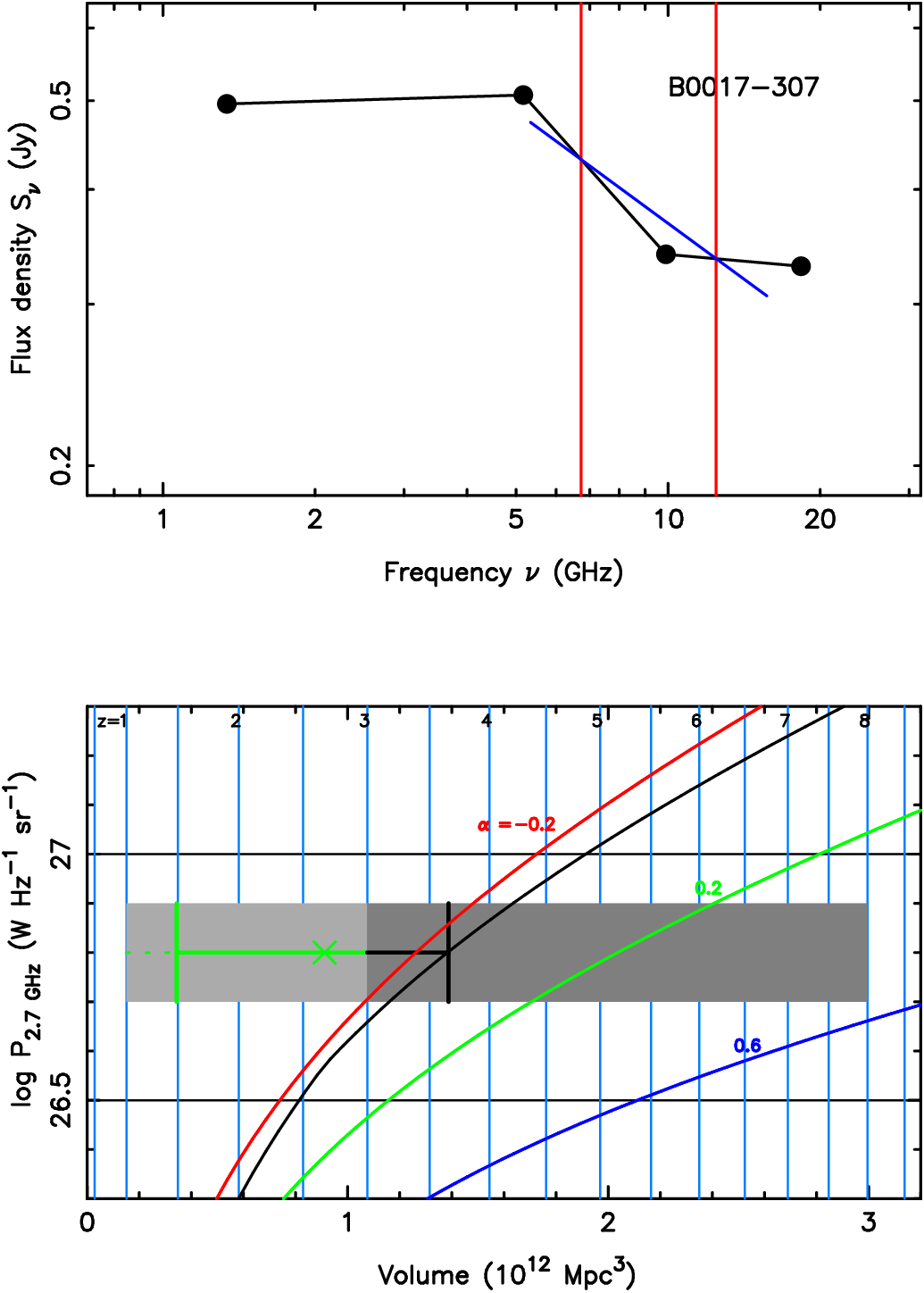}


\caption{Two sources to illustrate redshift limits in the single-source
analysis. For each object, the upper panel shows the spectrum in the rest
frame, while the lower panel shows the object in the $P-V$ plane. In each case
the individual cutoff lines are shown as the segmented black curves in the
$P-V$ plane, while the smooth coloured curves represent completeness limits at
$S_{\rm 2.7\,GHz} = 0.25$~Jy for sources whose spectra are described by single
power laws with indices as shown. Light grey bars represent the predictive
region $1 \le z \le 3$, dark grey the region $3 \le z \le 8$ in which the
object might be visible; the width of these bars is irrelevant. In the case of
PKS 2210-257, the individual survey limit suggests that the object should be
visible out to z=4.54. However, it is not visible to this redshift as an object
with an observed 2.7 - 5.0~GHz spectrum flatter than $-0.4$; the spectrum
(shown blue) steepens to an effective index of $\le -0.4$ at a redshift of
3.22, for which the `observed' frequencies of 2.7 and 5.0 GHz are shown in the
upper panel by the red vertical lines. The object therefore drops from the
sample at this redshift. PKS 0017-307 does not enter the observed region of the
diagram as an object with spectral index $\ge -0.4$ until a redshift of 1.48 is
reached. Again in its upper panel, the red vertical lines indicate rest
frequencies at 2.7 and 5.0 GHz for the critical redshift of 1.48, at which
point the spectrum `flattens' to have $\alpha \ge -0.4$, denoted by the blue
line. The object runs into its observable limit at z = 3.60 in the predictive
region as shown. The smaller accessible volume in the `observed' region results
in a scaling up of its prediction via the ratio of accessible volumes.}

\label{p-vdemo}

\end{figure*}
The second of these two effects is illustrated by PKS~2210-257 in the left-most
panel of Fig.~\ref{p-vdemo}. As the effective spectrum steepens with increasing
redshift, the accessible region in the redshift range $3<z<8$ is reduced
substantially from that given by the power cutoff line. The prediction from
such an object reduces correspondingly.

\item
There is a third effect due to spectral measurement affecting the `observed'
region $1<z<3$, if the spectrum steepens at frequencies below the survey
frequency. The effect is illustrated in the right panel of Fig.~\ref{p-vdemo}:
objects such as PKS~0017$-$307 are invisible to us as `flat-spectrum' sources
at redshifts below a cutoff point at which the upward curvature makes them
appear `steep-spectrum'. This affects only a few sources, but for these, it
means an increase in prediction contribution calculated as the ratio of
(accessible volume $3<z<8$)/(accessible volume $1<z<3$).

\end{enumerate}

When the available spectral measurements, relatively complete for Sample~2 at
five frequencies, are considered for each source, the prediction
(Table~\ref{results}) is 28.8 sources, differing now from the `observed' number
17.8 by just 2.9$\sigma$. The raised level of the power cutoff does most of the
damage. It is this use of the data for the Parkes 0.5-Jy sample which we
believe yields the relatively low level of significance for a redshift cutoff
found by \cite{jar00}.

However there is a fundamental problem with using the 8.40-GHz data. This can
be shown by using a set of 8.87-GHz flux densities measured in 1972
\citep{shi73}, roughly contemporaneous with the 2.7-GHz surveys.  There are 40
sources in Sample~2 with these `old' measurements, one of which, PKS 1532+016
at $S_{\rm 8.87\,GHz} = 1.16$ Jy, was not measured by \cite{wri90b}. There are
clearly large flux-density variations at 8~GHz, the wildest being for PKS
1402$-$012, 0.67 Jy in 1972, 0.15 Jy in 1989. If the 1972 8.87-GHz measurements
are used in preference to the 1989 8.40-GHz measurements, the prediction is
38.6 QSOs (57 contributors) in the range $3~<~z~<~8$. This is very
significantly higher than the prediction of 28.8 (53) sources using only
8.40-GHz data, exceeding the 17.8 sources `observed' by 4.9$\sigma$.

\begin{figure}[htb]

\vspace{8.0cm} \includegraphics{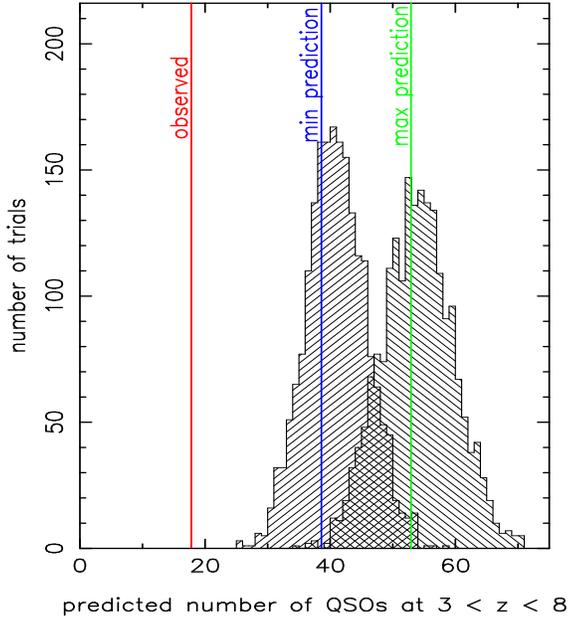}


\caption{The predicted numbers of QSOs at high redshifts. The histograms are
the results of 2000 trials of end-to-end bootstrap testing. The left-most
histogram was compiled using high-frequency spectral data at 8.40~GHz and
8.87~GHz, with the 8.87-GHz (near-contemporary) flux densities used in
preference if both were available. The right histogram represents 2000 results
from bootstrap testing when {\it only} the 8.87-GHz flux densities were used.
The vertical line to the left indicates the observed number of QSOs at high
redshifts, as discussed in the text. These bootstrap test were carried out in
the simple geometry $\Omega_{\rm m} = \Omega_{\rm tot} = 1$ but are closely
indicative of results for the $\Lambda$ - dominated geometry adopted here (see
Table~\ref{results} and text).}

\label{boot}

\end{figure}

The problem is a simple one. \emph{Measuring high-frequency flux densities some
time after the original survey gives a biased estimate of the spectrum.} Any
flux-limited survey preferentially selects variable sources in an up-state,
whereas flux-density measurements many years later reflect sources in a mean
state. The result is that the spectra are artificially steepened. In the
present case the result is an underestimate of numbers of objects predicted at
high redshifts. It is the variations at frequencies \emph{above} the survey
frequency which matter in this; variations at the lower frequencies are small
to insignificant in comparison.

The result emphasizes how responsive the predictions are to flux measurements,
and how crucial it is to use contemporaneous measurements. If this much change
comes about with replacing the 8-GHz flux densities of just 40 sources with
near-contemporary measurements, it is certain that the prediction of 38.6
sources based on using all the remaining (non-contemporary) 8.40-GHz flux
densities represents an underestimate or lower limit. If the 8.40-GHz flux
densities are ignored and only the 8.87-GHz data used as flux densities at
frequencies above 5.0 GHz, the result is a prediction of 50.5 sources. This
must be an overestimate. The 8.87-GHz flux densities were measured
preferentially for bright sources at high frequencies, and thus favour objects
well above survey-limit lines. We conclude that on the hypothesis of uniform
space distribution, somewhere between 38.6 and 50.5 sources are predicted to
have redshifts between 3 and 8 for Sample~2.

To assess the uncertainties  an end-to-end bootstrap experiment was run for the
two possibilities: (i) using as high frequency data only the 8.87-GHz (1972)
measurements, and (ii) using the combination of 8.87 and 8.40-GHz measurements,
with the former taking precedence if measurements at both flux densities were
available. Because of computing time constraints we had to run this experiment
using the simple geometry of\ $\Omega_{\rm tot} = \Omega_{\rm m} = 1$. However
as Table~\ref{results} shows, the predictions in this geometry are very similar
to the predictions of the $\Lambda$ - dominated cosmology, the numbers in
question being 38.6 and 52.9 for the simple geometry vs. 38.6 and 50.5 for the
$\Lambda$ - dominated geometry. The uncertainties should be representative. In
the bootstrap sampling, random redshifts were assigned to each source from the
total sample of redshifts. The flux densities for the source were then
`corrected' to that particular redshift making use of the measured redshift of
the object. The results are shown in Fig.~\ref{boot}. Some 2000 trials for each
of the two possibilities produced no prediction as low as the `observed' number
of 17.8 QSOs at $z > 3$.

There are two results from this analysis:
\begin{enumerate}
\item
The true prediction of numbers of QSOs at $3<z<8$ for a uniformly-filled
universe lies between 38.6 and 50.5 objects for Sample~2. This is to be
compared to the `observed' number of 17.8. An end-to-end bootstrap test
indicates that the difference is highly significant.
\item
High-frequency flux-density measurements that are non-contemporary are
dangerous. They bias the spectral statistics of any variable-flux sample,
because surveys pick out variable sources in their high state, and not in their
average state.

\end{enumerate}

\section{The form of the evolution}

\label{evol}

Figs.~\ref{rlf_dif} and \ref{rlf_z} indicate that the form of the evolution,
and in particular the shape of the decline at high redshifts,  cannot be
inferred directly. As an indirect route, we used Sample~2 and proceeded as
follows:

\begin{figure}

\vspace{13.0cm} \includegraphics{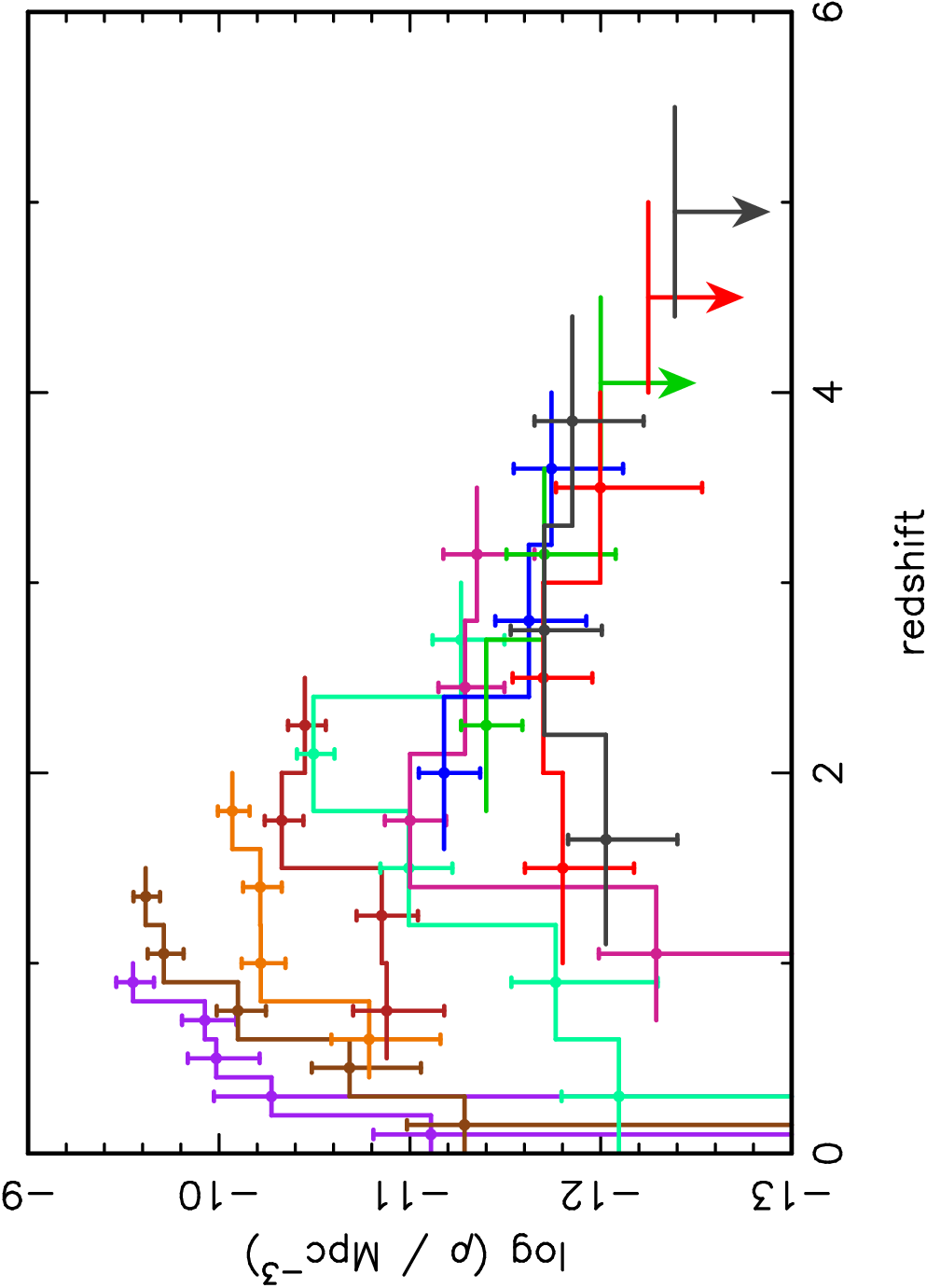} \includegraphics{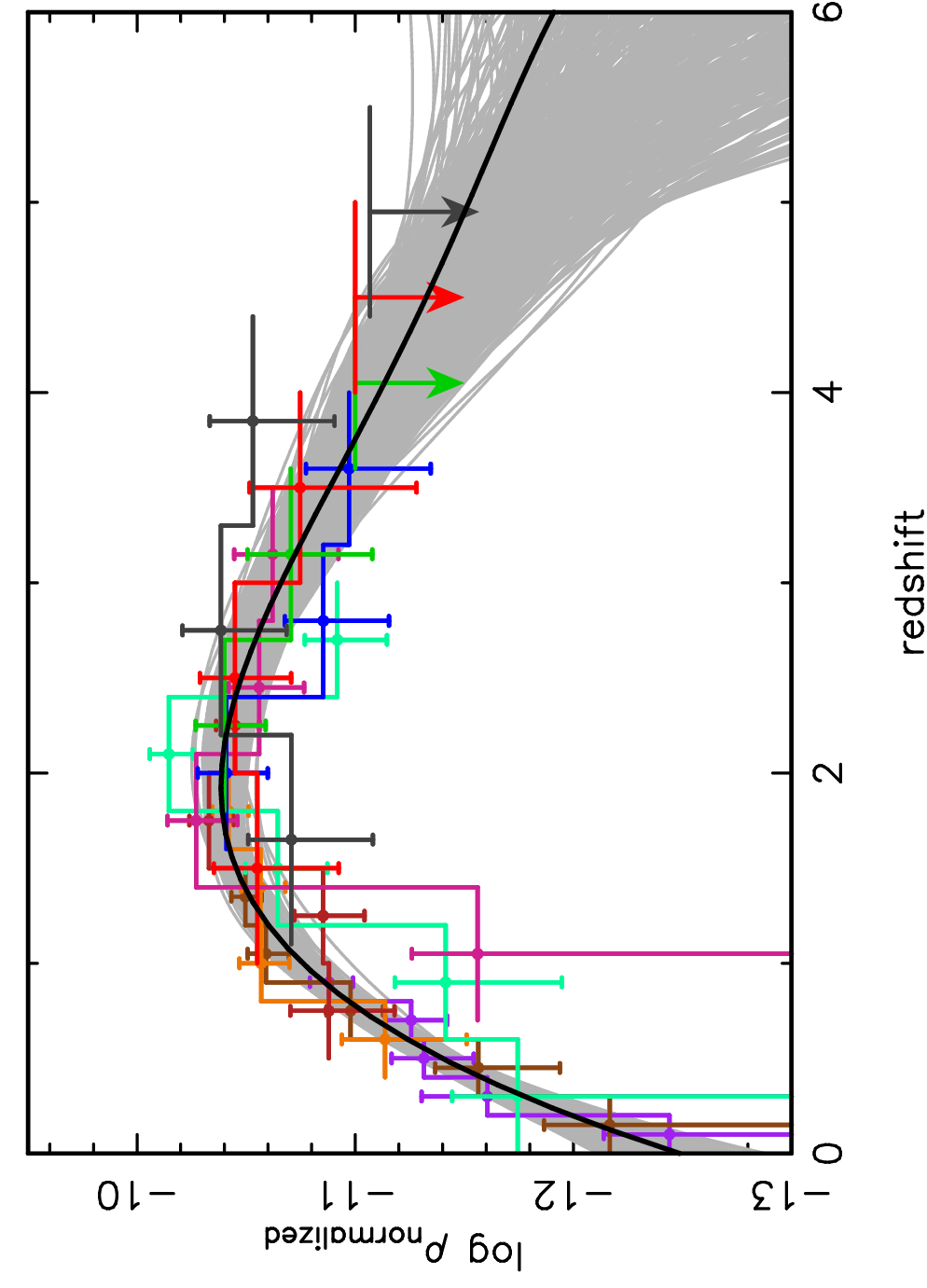}


\caption{Above: space density $\rho$ vs redshift. The individual RLFs are each
complete from $z=1.0$ to $z=5.5$ in steps of $\Delta z = 0.5$, in the order
purple, brown, orange, dark red, light blue, turquoise, blue, green, red, grey.
Below: these RLFs normalized to agree over the range at $1.0< z <2.5$. The bold
black line is a least-squares fit with a polynomial of fifth order, given in
the text. The grey lines represent 1000 bootstrap trials. In this process, fits
which resulted in lines of positive slope beyond $z=5$ were rejected.}

\label{rlf_zc}

\end{figure}

\begin{enumerate}
\item Ten redshift limits were set up, from $z_l=1.0$ to $z_l=5.5$ in steps of $\Delta z = 0.5$.
We then determined the
 combination of $P_{2.7}$ and effective spectral index yielding the maximum number of
contributors to the RLF for the sub-sample complete to each of these redshift
limits. The numbers of RLF contributors in these complete sub-samples ranged
from 285 at $z=1.0$ down to just 9 at $z=5.5$; the numbers decrease since
higher luminosities are needed to be complete to the larger redshift limits.

\item For each of these 10 sub-samples, we computed the RLF using the $1/V_{\rm max}$
contributions already calculated for previous estimates of luminosity
functions. For the two samples with redshift limits at 5.0 and 5.5, there are
no sources in the upper bins, and to use this observation we assigned an upper
limit of one source to each of them. The results are shown in
Figure~\ref{rlf_zc}, upper panel.

\item Although these RLFs are each now complete from $z = 0$ to $z=z_l$,
individually they are inadequate to trace the whole QSO epoch. Those with
smaller values of $z_l$ by definition cannot reach large redshifts, while those
with the larger $z_l$ are severely noise limited, particularly at the
low-redshift ends. To combine these results to define the overall space
behaviour, we normalized each curve to agree statistically over the range $1.0
< z < 2.5$. We then fitted a least-squares polynomial through the points, with
the results shown in the lower panel of Fig.~\ref{rlf_zc}. The heavy black line
in this diagram is
$$
{\rm log} \rho = -12.49+2.704z-1.145z^2+0.1796z^3-0.01019z^4.
$$

\item Finally a constrained bootstrap experiment of 1000 runs was used to give
an approximation to the uncertainty; the grey area in the figure is the result.

\end{enumerate}

Normalizing as described is only valid if the form of evolution is independent
of radio luminosity. There is some indication in the data (Fig.~\ref{rlf_zc},
upper panel) that the turnovers set in at redshifts increasing with luminosity.
If so, then normalizing as described would broaden the maximum, and the overall
curve would be representative of the overall evolution form, but in no sense
formally accurate. Moreover, the complete individual pieces of luminosity
function are not statistically independent, so that the shaded area is
indicative of uncertainty, but again is not accurate in a formal sense.

\begin{figure*}


\vspace{11cm}

\includegraphics{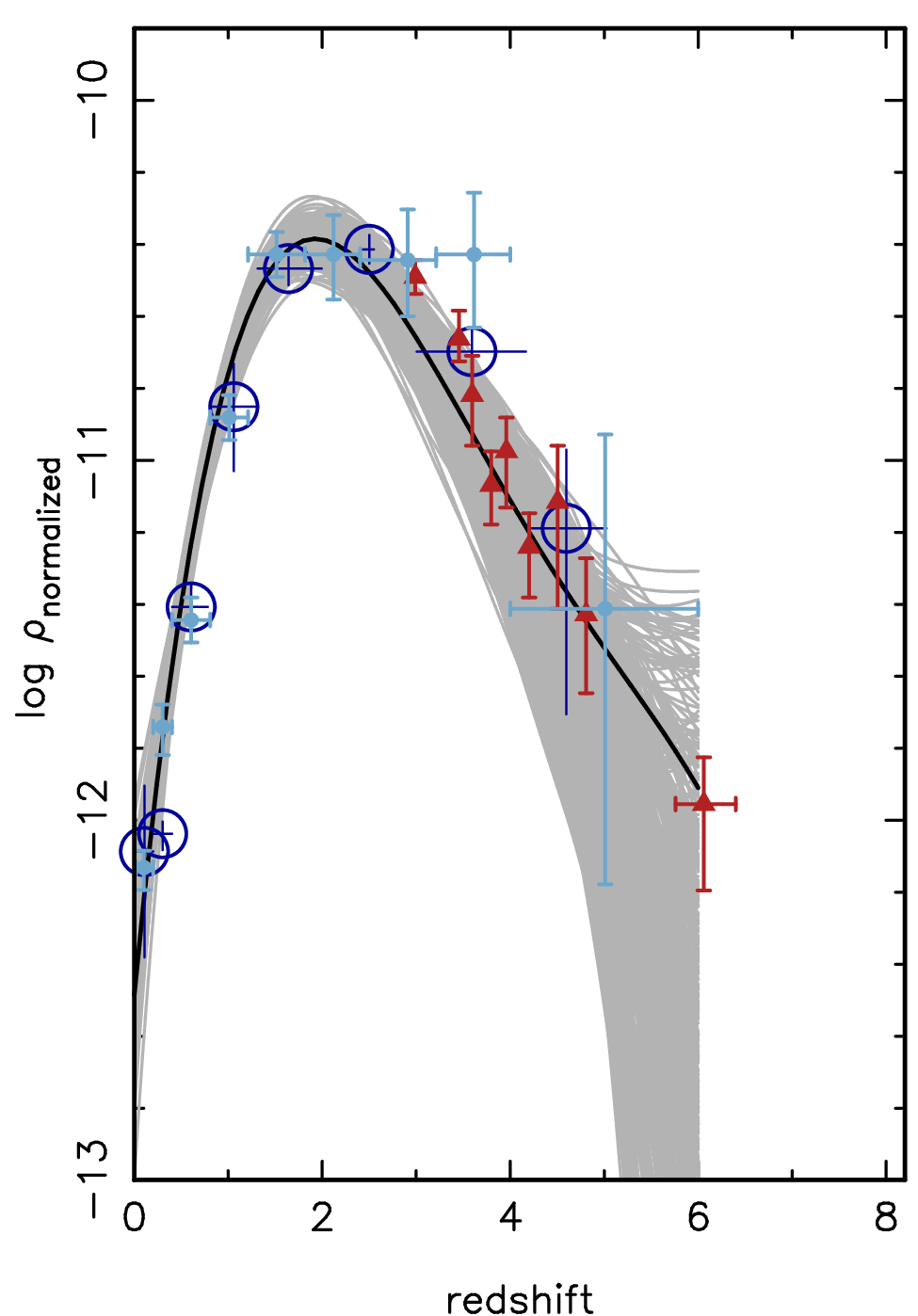}

\includegraphics{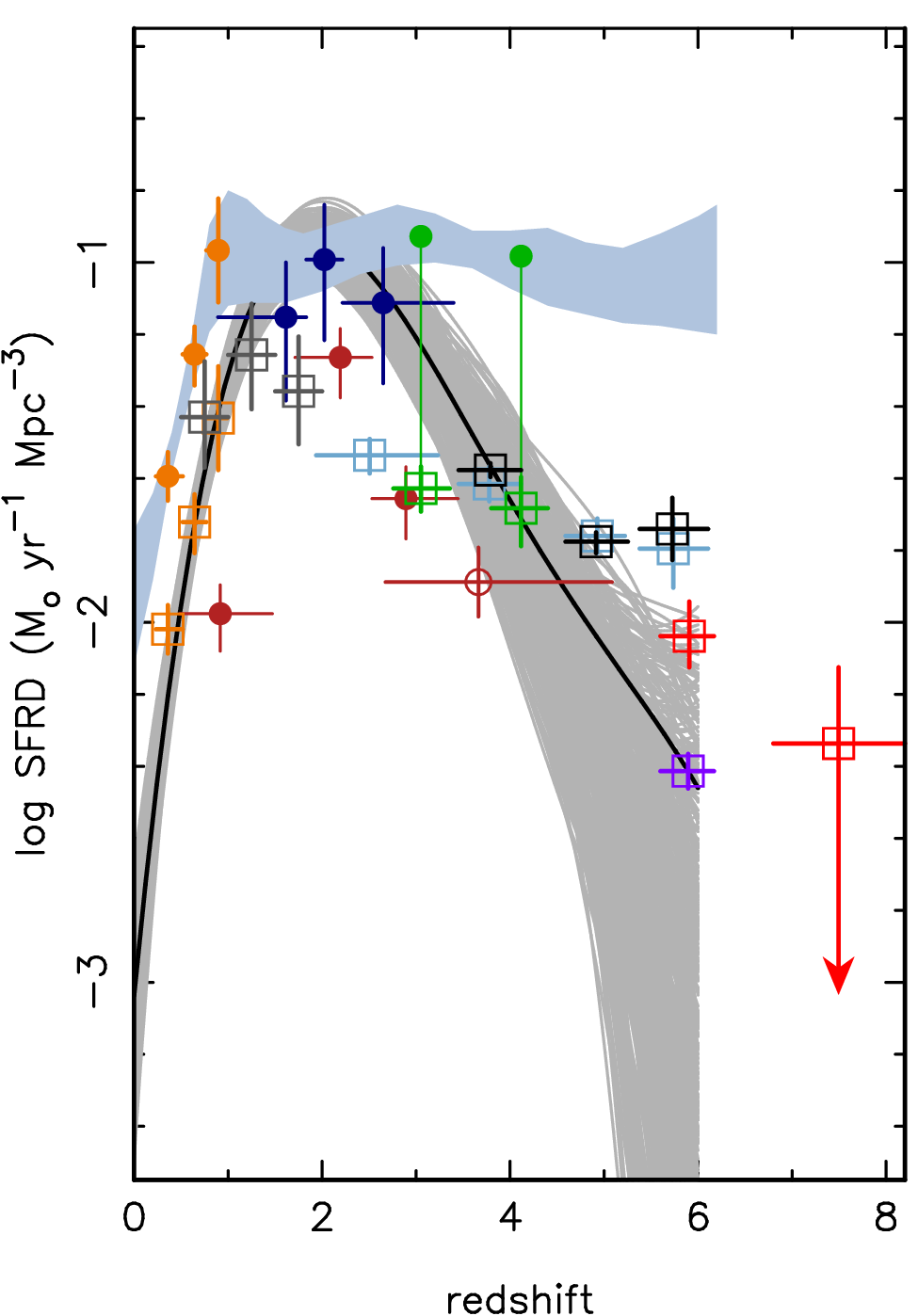}

 \caption{Left: Relative space density of QSOs ($\rho$)
as a function of redshift.  The shaded area and black line represent the
current QSO space-density determination from Figure~\ref{rlf_zc}. Light blue
filled circles show the soft X-ray data from Chandra and XMM-Newton surveys
\citep{has04}, while the dark blue open circles show the results from combined
Chandra and ROSAT surveys \citep{sil04}. Space density behaviour of
optically-selected QSOs is given by the set of dark red triangles; the data are
from \cite{sch95}, \cite{fan01c} and \cite{fan04}. The point at $z\sim6$ is
taken from \cite{fan04} and due to conversion between geometries there is
uncertainty in the ordinate of $0.1$. The X-ray and optical QSO data were
scaled vertically to match the current determination of space density at
redshifts 2 to 2.5. Right: Star formation rate density (SFRD; units adopted by
\citealt{bla02}), with the shaded area and black line again showing the current
estimate of radio QSO space density. Data in optical and near-IR bands are
distinguished as squares: orange \citep{lil96}, grey \citep{con97}, green
\citep{ste99}, light blue \citep{bou04a}, black \citep{gia04}, purple
\citep{bun04} and red \citep{bou04}. In general these data have not been
corrected for extinction. Green dots show results of the extinction correction
of 4.7 suggested by \cite{ste99} for their data; the light blue band represents
an estimate of the SFR determined from these and other points as analyzed by
\cite{bou04}. Measurements from Far-IR and sub-mm observations are shown as
circles: FIR as orange filled circles from \cite{flo99}, and sub-mm as dark red
filled and open circles and blue filled circles from \cite{bou04}, discussed in
the text. The current space density determination (black curve) was scaled to
match the \cite{lil96} points, the orange open squares.}

\label{rlf_sfr}

\end{figure*}

These results enable comparison with other high-redshift observations.
Fig.~\ref{rlf_sfr} shows the shaded area of Fig.~\ref{rlf_zc} in the
background, with data from recent compilations of space density as a function
of redshift for AGNs selected at X-ray and optical wavelengths (left panel),
and star-formation rate (SFR) as a function of redshift (right panel).

Agreement with the form of the X-ray-selected QSO evolution is remarkably good.
\cite{sil04} found the X-ray decline to agree in form with the optical decline
determined by Fan and co-workers \citep{fan01a,fan01b,fan01c} from SDSS.
Silverman et al. also showed that the COMBO-17 survey results of \cite{wol03}
follow the X-ray data closely. The \cite{has04} X-ray AGN results are again in
very good agreement with the current determination. There is thus general
accord between the dependence of space density for QSOs found at radio, optical
and X-ray wavelengths, all showing a rapid rise in co-moving space density to
$z \sim 1.5$ followed by declining space densities at $z > 3$. However, there
are strong dependencies of evolution form on luminosity, certainly for the
optical and X-ray samples as noted by \cite{has04} and \cite{sil04}; and there
may be such dependence for the current radio-selected sample. The dependence on
luminosity is well illustrated by the fact that the rising curve of
(lower-optical-luminosity) QSOs selected from the 2dF survey \citep{cro04} is
displaced to higher redshifts than the X-ray or radio-selected QSOs. The
current agreements are illustrative only; analysis of the significance must
await larger samples providing better definition of space density evolution as
a function of luminosity in each wavelength band.

The relation between QSO space evolution and star-formation-rate history is not
so clear. The general similarity was first noted in 1997 \citep{wal98}; the
rise to redshifts of $\sim1.5$ appears to be of the same form. But at this
redshift it appears that determinations of the SFR from UV, optical and near-IR
measures produce a different form of epoch dependence, with an abrupt
transition to a law almost constant (or diminishing gradually) with increasing
redshift out to $z > 6$. There are substantial uncertainties in what extinction
correction to apply; but this form appears to hold whether or not the data are
extinction-corrected (provided as claimed that the correction is not strongly
dependent on redshift). The open squares of Fig.~\ref{rlf_sfr}, uncorrected for
extinction, show the gradual decline, while the band \citep{cha04},
representing a fit to extinction-corrected data, shows a star formation rate
essentially independent of redshift to $z>6$. The data from sub-mm observations
(dark blue and red circles, filled and open; \citealt{cha04}) represent
estimates from radio-identified sub-mm galaxies (red circles), and from these
galaxies and sub-mm galaxies combined (blue circles). Chapman et al. point out
that the similarity of star-formation-rate contributions at $z\sim 2$ suggests
that the total SFR from all populations may exceed the current estimates
significantly.

It remains somewhat puzzling that the sub-mm galaxy star-formation-rate appears
to drop beyond $z = 3$, and that it therefore resembles the AGN space-density
law rather than that of the galaxies detected in the optical and near-IR. This
may be superficial, in that the points are lower limits, and additional
components may be found. It is perhaps less puzzling that the AGN space-density
law differs from the overall SFRD in the sense observed. On current
hierarchical pictures, redshifts beyond 3 represent the era of rapid galaxy
assembly; and there may be a delay before the large enough galaxies have
developed to host massive black holes, or before the galaxy-building process
provides orbit organization appropriate to fuel such black holes.

\section{Summary}

\label{sum}

We summarize the results of this and the preceding two papers
\citep{jac02,hoo03}.

\begin{enumerate}

\item (Paper~1) The initial goal of the project was to search for high-redshift
QSOs without dust bias. Optical counterparts for essentially all flat-spectrum
objects in the sample were obtained. No QSOs at redshifts greater than 5 were
found.

\item At the fainter flux densities ($S_{2.7} < 0.4$ Jy) and optical magnitudes
($B_J > 20.0$), substantial numbers of  flat-spectrum radio galaxies are
present (Paper~1, Fig.~7). These may have influenced previous claims for a
hidden population of heavily-reddened QSOs; `red' QSOs \citep{rie79} do not
appear to constitute a major fraction of the total population sampled here.
Amongst the reddest of the stellar identifications, about one-quarter are
BL\,Lac objects, compared to 9~per~cent with no colour selection, supporting
the synchrotron interpretation for many of the red objects originally proposed
by \cite{rie79}. However, the discovery of molecular line emission from a
sample of the redder QSOs by \cite{car98} indicates the likely presence of dust
in some systems, possibly concentrated in dense nuclear tori.

\item (Paper~2) A composite optical spectrum for flat-spectrum QSOs derived from the
present sample shows clear qualitative differences in comparison with
radio-quiet composite spectra: the Ly-$\alpha$(1216\AA) and CIV(1549\AA) lines
are stronger in the current sample. There appears to be no significant
difference in the Ly-$\alpha$ decrement between radio-loud and radio-quiet
QSOs.

\item (Present paper) The redshift distribution has been derived for the
quarter-Jansky sample, with consideration given to the different constituents -
red galaxies, BL\,Lac objects and QSOs. Comparison with the space-density
models of \cite{dun90} and \cite{jac99} indicates substantial agreement,
although refinement of space-density modelling is clearly possible using the
current data.

\item Radio luminosity functions for flat-spectrum QSOs have been
calculated using the $1/V_{\rm obs}$ method. These show the rapid increase in
space density between $z = 0$ and $z = 1$, a flattening between $z = 1$ and $z
= 2.5$, and evidence of declining space density to yet higher redshifts.
Attention is drawn to a selection effect producing an apparent flattening of
the luminosity functions towards lower powers in each redshift shell. This
effect, due to intrinsic spread in radio spectra, may have gone unrecognized in
previous analyses, leading to overinterpretation of similarities in form of the
RLF at each redshift.

\item The reality of a redshift cutoff has been verified using the
`single-source survey' $V_{\rm max}$ method, in which each QSO in the sample is
used to predict the number of sources observable at higher redshifts on the
assumption of uniform space density. The technique is robust and model-free,
and it makes use of all QSOs in the sample, rather than limiting the statistics
to the few objects of highest radio luminosity.

\item The result of this analysis has been to demonstrate that a space-density
diminution exists at $z > 3$, at significance levels of $>4\sigma$. Precise
significance cannot be evaluated, because it is shown that spectral data at
high frequencies are critical, and that non-contemporary flux-density
measurements seriously bias the results towards reducing the apparent redshift
diminution.

\item In the light of this high-redshift diminution, an estimate has been made
of the overall evolution for the radio-loud QSO population. The model-free
approach has been retained, with the form mapped simply by fitting a
least-squares polynomial to the data. This was followed by a bootstrap analysis
- 1000 more polynomial fits - to provide an indication of the uncertainty.

\item The form of this evolution was compared with that determined for
X-ray QSOs (Chandra, XMM-Newton and ROSAT results) and for optically-selected
QSOs (primarily SDSS results). Agreement is excellent. It was further compared
with the evolution of star formation rate. Current best estimates of this from
optical and near-IR data show that although the initial rise may be of similar
form, there is divergence at redshifts beyond 1.5. While the AGN space density
dives down at $z>3$, the SFR appears shows relatively little dependence on
redshift to $z>6$. The picture from sub-mm measurements is less clear.
Uncertainties are large, but the SFR determined from sub-mm data appears
curiously to follow the AGN relation more closely than it follows the SFR law
found from galaxies detected with optical and near-IR measurements.

\item In view of the diminution in space density to high redshifts, radio-loud
QSOs would appear to have little role to play in the reionization epoch
$6<z<17$.

\end{enumerate}

The reality of the redshift cutoff for radio-selected QSOs and its similarity
with that observed for optically-selected QSOs leads to the conclusion that the
apparent cutoff for the latter is real and is not due to dust. This conclusion
is consistent with the results from the CORALS survey \citep{ell01,ell02}, in
which a complete sub-sample of radio-selected QSOs from the Parkes
quarter-Jansky sample was searched for damped Ly-$\alpha$ (DLA) systems. Little
significant difference for the comoving mass density of neutral gas was found
between the estimate from the CORALS sample and those from previous (optically
selected) samples. No major population of high-column-density absorbers has
been missed. Our view of the high-redshift diminution of the QSO population
does not appear to be dominated either by intrinsically dusty systems or by a
major Universal dust screen.

\noindent{\it Acknowledgements.} We are very grateful to Matt Jarvis and Steve
Rawlings for a helpful dispute. We thank Greg Bryan, Xiaohui Fan, Charles
Jenkins, Anna Sajina and Douglas Scott for valuable discussions.  We are
grateful to G\"{u}nther Hasinger and colleagues for providing us with data and
results prior to publication. The National Radio Astronomy Observatory is
operated by Associated Universities, Inc. under a cooperative agreement with
the U.S. National Science Foundation.





\end{document}